\documentclass[12pt]{article}
\usepackage[utf8]{inputenc}
\usepackage[T1]{fontenc}
\usepackage[margin=1.25in]{geometry}
\usepackage{mathtools}
\usepackage{amsmath}
\usepackage{amsfonts}
\usepackage{amssymb}
\usepackage{amsthm}
\usepackage{natbib}
\usepackage{setspace}
\usepackage{graphicx}
\usepackage{bbm}
\usepackage{booktabs}
\usepackage{hyperref}
\hypersetup{hidelinks}
\hypersetup{
    colorlinks=true,
    citecolor=blue,
    linkcolor=blue,
    pdfborder={0 0 0}
}

\usepackage{caption}
\usepackage{subcaption}
\usepackage{pgfplots}
\pgfplotsset{compat=1.18}
\usepackage{tikz}
\usepackage{float}

\newtheorem{theorem}{Theorem}
\newtheorem{proposition}{Proposition}
\newtheorem{lemma}{Lemma}
\newtheorem{corollary}{Corollary}
\newtheorem{definition}{Definition}
\newtheorem{assumption}{Assumption}
\newtheorem{lemmaS}{Lemma}

\theoremstyle{remark}

\newtheorem{example}{Example}

\renewcommand{\hat}{\widehat}
\renewcommand{\bar}{\overline}
\DeclareMathOperator{\marg}{marg}

\newcommand{\one}{\mathbbm{1}}
\newcommand{\E}{\mathbb{E}}
\newcommand{\cx}{\preceq_{\mathrm{cx}}}

\title{\textbf{Coarse Revenue Guarantee in First Price Auction and Beyond}\thanks{I am indebted to Songzi Du for his invaluable guidance on this project. I am immensely grateful to Aram Grigoryan, Jun Liu, Joel Sobel, and Emanuel Vespa for their useful suggestions for this paper. I also thank Cuimin Ba, Arjada Bardhi, Aislinn Bohren, Benjamin Brooks, Daniel Gottlieb, Duarte Gonçalves, Menglong Guan, Jiangtao Li, Markus Moller, Joao Ramos, Doron Ravid, Denis Shishkin, Vasiliki Skreta, Alex Smolin, Joel Watson, Frank Yang, and Mengxi Zhang, as well as the audiences at EC~'24, GAIMSS~'24, Stony Brook Game Theory Conference 2024, and UCSD TBE Workshop for valuable conversations and constructive comments. A previous version of this paper was circulated as “Revenue Guarantee across Bayes Coarse Correlated Equilibria.” Claude was used for assistance with simulation, proof verification, and text refinement. All errors are my own.}}

\author{Linchen Zhang\thanks{Department of Economics, University of California San Diego. Email: \texttt{liz032@ucsd.edu}.}}
\date{\today\thanks{First version: January 2024.}}

\begin{document}

\singlespacing

\maketitle

\begin{abstract}
\noindent I study the coarse revenue guarantee (CRG) of the single-unit first-price auction with common values. The CRG is the minimum expected revenue over all information structures and all Bayes coarse correlated equilibria (BCCE), the equilibrium notion to which no-regret learning leads. Identical play, in which all bidders submit the same bid, attains the worst-case revenue, and this reduction yields a closed-form characterization. The coarse guarantee is positive and increases with the number of bidders $N$. It lies strictly below the revenue guarantee of \citet{BBM2017} at every finite $N$, and the gap between the two closes at rate $\log N/N$. I extend the analysis to reserve prices, symmetric and asymmetric priors, other standard auctions, and an unknown value distribution.
\end{abstract}

\newpage

\onehalfspacing

\section{Introduction}

The robustness of mechanisms to the information structure is a central question in mechanism design. A seller who knows the distribution of the good's value but not the buyers' information structure may want a lower bound on revenue that holds across every environment consistent with that distribution. \citet{BBM2017} provide such a bound for the first-price auction. They characterize the \textbf{revenue guarantee} (RG), defined as the minimum expected revenue across all information structures and all Bayes correlated equilibria (BCE), and show that it is positive and stays bounded away from the total surplus even as the number of buyers grows large. The BCE framework assumes fully Bayesian rationality. In a BCE, a bidder who is told to submit $a_i$ after observing signal $s_i$ must be unwilling to change that bid \emph{at that signal}. Obedience holds signal by signal, which requires each bidder to process their private information and best-respond conditional on it. This is the natural requirement when bidders are sophisticated Bayesian agents.

But it may not be the appropriate requirement when bidders are algorithms. In modern electronic marketplaces, and increasingly in advertising exchanges and high-frequency trading venues, bids are submitted by automated agents. These agents adjust their bids over time by simple learning rules rather than by solving a Bayesian inference problem. The standard performance criterion for such rules is \emph{no regret}: the algorithm guarantees that, in hindsight, no single fixed action played throughout the history would have earned more on average. That is an unconditional guarantee. It says nothing about whether the algorithm's behavior would have been optimal in any particular state, and a no-regret algorithm can have large regret conditional on individual states without ever violating its guarantee.\footnote{See \citet{StructuralAI} for an econometric framework built on this weaker restriction, and \citet{Nekipelov2015} for inference from no-regret bidders.}

The solution concept that addresses this weaker notion of rationality is Bayes coarse correlated equilibrium (BCCE), in which each player must be unwilling to switch to a single fixed action played at every signal. When regret is measured against such fixed actions, the empirical distributions generated by no-regret dynamics in a Bayesian game accumulate on the set of BCCE, not on the smaller set of BCE. \citet{Hartline2015} prove convergence of this kind for independent private values, with regret measured separately for each value of a player, so that a deviation may depend on that value. A seller who faces algorithmic bidders and wants a lower bound on revenue that holds for every outcome to which those algorithms can converge must therefore minimize over BCCE. This paper carries out an exercise similar to that of \citet{BBM2017} for the first-price auction, with BCCE in place of BCE.

I define the \textbf{coarse revenue guarantee} (CRG) as the minimum expected revenue across all information structures and all BCCE, and I ask four questions. How large is it in the first-price auction, and which equilibrium attains it? How much does the seller lose by facing coarsely rational bidders rather than interim rational ones?\footnote{The word \emph{interim} refers to timing and is used throughout as the counterpart of \emph{coarse}. In a Bayes correlated equilibrium a bidder decides whether to deviate after seeing their information, so the constraint is checked at the interim stage, conditional on the signal and the recommended bid. A coarse deviation, by contrast, is one fixed bid chosen before any information arrives, and its constraint averages over all states, which makes it an ex ante requirement. Section \ref{sec: model} gives the formal definitions.} What can the seller do to raise it? And how do the results change across settings, namely symmetric and asymmetric priors, other auction formats, and an unknown value distribution?

\paragraph{Characterization through identical play}
The first contribution is a tractable characterization built on a structural result. In the first-price auction, from any BCCE one can construct another BCCE with weakly lower revenue in which every buyer submits the same bid with probability one. The worst case can then be taken to be \emph{identical play}: bidders do not use their private information and all submit the same bid. The construction, which I call pooling, works because two effects act in the same direction. First, revenue in a first-price auction is the highest bid, so replacing a dispersed bid profile with a common bid that is no higher than the highest bid in the profile lowers revenue. This effect holds under any notion of rationality. Second, the replacement can be arranged so that the joint distribution of the value and the \emph{highest competing bid} faced by a randomly chosen buyer is unchanged, and this is what preserves the equilibrium constraints. A coarse deviation is a single fixed bid, and the profitability of a fixed bid depends on one's rivals only through the best bid among them. Holding that joint distribution fixed while lowering the price leaves the buyers' average payoff from deviating unchanged and raises their average payoff from obeying, and under identical play every buyer receives the average.

The structure of this worst case is where the coarse analysis departs from the BCE benchmark. In \citet{BBM2017}, the worst case is supported by an information structure in which bidders learn something about their rank and play symmetrically \emph{conditional} on that information. Under coarse rationality, a worst case is instead unconditional identical play. Bidders who cannot condition a deviation on their signal can be induced to submit a common bid, so that no bidder acts on information the others lack.

With $N$ bidders, the reduction also turns an $N$-dimensional problem into a one-dimensional one. The seller's worst case no longer requires a search over joint distributions of values, signals, and $N$-dimensional bid profiles, since it is determined by the joint distribution of the value $v$ and a single common bid $c$. Theorems \ref{thm:continuous_char} and \ref{thm:general} solve this problem in closed form for an arbitrary value distribution. In the worst case, buyers bid zero when the value is below a threshold, so the good is sold at a price of zero, and above the threshold they submit a common bid that increases with the value. The surplus from the zero-bid region is what keeps buyers from deviating to a slightly higher bid. The guarantee is strictly positive for every non-degenerate prior, that is, one not concentrated on a single value, and strictly increasing in the number of buyers.

\paragraph{The cost of coarseness}
Because every BCE is a BCCE, the coarse guarantee is weakly below the interim guarantee of \citet{BBM2017}. Proposition \ref{prop:strict} shows that the inequality is strict at every finite $N$, so no BCE attains the coarse minimum. One way to see why is that the coarse worst case has a structure that interim obedience rules out: a bidder who is told to bid zero knows that the value is low and would gain by bidding slightly more. The gap is large when the number of bidders is small.

Nevertheless, Theorem \ref{thm:asymptotic} shows that this gap closes as $N$ grows. The coarse constraint is satisfied by lowering revenue until each buyer's equilibrium payoff is at least the payoff from every fixed deviation, and with $N$ buyers sharing the surplus, that payoff is only $1/N$ of the total. As $N$ grows, the additional revenue reduction that the coarse constraint permits goes to zero, and the two guarantees coincide in the limit. Proposition \ref{prop:rate} shows, however, that the coarse guarantee converges more slowly. It reaches the common limit only at rate $\log N/N$, compared with $1/N$ for the interim guarantee, so the cost of coarseness with $N$ buyers, $RG_N-CRG_N$, also vanishes at rate $\log N/N$.

\paragraph{What the seller could do}
The next question is what the seller can do to increase the coarse guarantee. One natural instrument is a reserve price, under which the good is withheld if the highest bid is below the reserve. Section \ref{sec: design} shows that the identical play reduction still holds with a reserve price once the pooling construction is modified, so the same tools apply and the guarantee can be characterized for every reserve below a bound given in Theorem \ref{thm:reserve}. Proposition \ref{prop:optimal_reserve} then gives a necessary condition for a reserve in this range that maximizes the guarantee. At such a reserve, the guarantee is attained by a worst case in which buyers with low values stay out and the others bid above the reserve, so that no buyer pools at the reserve and pays only that price.

\paragraph{Symmetric and asymmetric priors}
The main analysis assumes pure common values. The interim characterization of \citet{BBM2017} covers symmetric priors with arbitrary correlation, so it is natural to ask what changes when values may differ across buyers, and Section \ref{sec: extension} takes up this question. The pooling argument behind the identical play reduction then does not extend. When values differ, the adversary who aims to reduce the seller's revenue has an additional instrument. Through an arbitrarily small bid premium, it can keep the price common and still allocate the good to the buyer with the highest value, and the obedience constraints continue to hold. Nevertheless, Theorem \ref{thm:symmetric} shows that the guarantee for any exchangeable prior, one that is unchanged when the buyers are relabeled, is bounded below by the guarantee of Theorem \ref{thm:continuous_char} with the mean value replaced by the expected highest value. The bound depends on the prior only through each buyer's own marginal and the expected highest value, so it extends to asymmetric priors with a separate set of constraints for each buyer, and Theorem \ref{thm:asymmetric} gives the resulting lower bound.

\paragraph{Ranking and equivalence across auctions}
Section \ref{sec: RGRanking} compares the first-price auction with the other standard auctions, following \citet{BBM2019RGE}. The comparison rests on one observation. Under identical play, every standard auction allocates by the same tie-breaking rule and charges the same price, so the outcome that attains the first-price guarantee is available in every format, and formats differ only in what a single deviator pays after outbidding the others. If the deviator pays their own bid, the identical play equilibria of the first-price auction are equilibria of the format. Its guarantee restricted to identical play is then the first-price guarantee, and its full guarantee is at most that. When the deviator pays the others' bid instead, as in the second-price and English auctions, outbidding them does not raise the price paid, pooling at a low price cannot be sustained, and an asymmetric equilibrium yields a guarantee of zero. The first-price auction has the highest guarantee in both cases, so a seller designing for no-regret bidders is led to the same format as one designing for Bayesian bidders.

\paragraph{Unknown value distribution}
Finally, Section \ref{sec: binary restriction} considers a seller who knows only the mean of the value distribution and an upper bound $\bar v$ on the value, and who wants a lower bound on revenue over BCCE. I first show that the coarse revenue guarantee is monotone in the convex order: a mean-preserving spread of a prior is weakly worse for the seller. Because the binary distribution on $\{0,\bar v\}$ is the most spread-out distribution on $[0,\bar v]$ with a given mean, it is the worst-case prior, and its guarantee is a lower bound that holds for every prior with that mean. I then solve for this bound in closed form.

The remainder of the paper is organized as follows. Section \ref{sec: lit} discusses the related literature, and Section \ref{sec: model} introduces the model and the two equilibrium notions. Section \ref{sec: identical play} establishes the identical play reduction, which Section \ref{sec: general characterization} uses to characterize the coarse revenue guarantee, its limit, and its rate of convergence. Section \ref{sec: design} studies reserve prices. Section \ref{sec: extension} extends the lower bound to symmetric and asymmetric priors, Section \ref{sec: RGRanking} compares the guarantee across standard auctions, and Section \ref{sec: binary restriction} treats an unknown value distribution. Section \ref{sec: conclusion} concludes. Omitted proofs are in the Appendix and the Supplementary Appendix.

\section{Related Literature} \label{sec: lit}

This paper contributes to three strands of literature: first-price auctions, robust revenue guarantees, and the connection between Bayes coarse correlated equilibrium and no-regret learning.

First, this work builds on the literature on the first-price auction. Earlier studies, such as \citet{Vickrey1961}, \citet{Milgram1982}, \citet{BATTIGALLI2003}, \citet{DEKEL2003}, \citet{FANG2006}, and \citet{Azacis2015}, relied on specific assumptions about the information structure. More recently, \citet{BBM2017} allowed for arbitrary information structures but restricted attention to Bayes correlated equilibrium. This paper characterizes the coarse revenue guarantee under the weaker notion of BCCE and develops coarse counterparts of the ranking and equivalence results of \citet{BBM2019RGE}. The closest earlier work is \citet{Feldman2016}, who study correlated and coarse equilibria of single-item auctions. They work in the complete-information model. I minimize revenue over all information structures and characterize the minimizing equilibrium exactly.

Second, this paper contributes to the literature on revenue guarantees, including \citet{BBM2017}, \citet{BBM2019RGE}, \citet{Du2018}, and \citet{Brooks2021}, by studying these guarantees beyond BCE. Work on the price of anarchy \citep{Tardos2013,Jin2023} gives related guarantees for welfare. This paper is concerned with revenue.

Lastly, this work contributes to the literature on BCCE and learning in games. Coarse correlated equilibrium was introduced by \citet{Moulin1978}. It has received relatively little attention in economics, perhaps because its set of deviations is small. The use of no-regret algorithms as bidders has renewed interest in it. In games of complete information, regret matching leads to correlated equilibrium \citep{Hart2000}, and \citet{Blum2008} bound the welfare of play in which every player minimizes regret. \citet{Hartline2015} show that in Bayesian games with independent private values, no-regret dynamics converge to a set of Bayes coarse correlated equilibria in which a deviation may depend on the player's own value. A growing literature studies auctions in which some or all participants are learning algorithms: \citet{Nekipelov2015} on inference from no-regret bidders, \citet{Braverman2018} on a seller's ability to exploit them, \citet{Deng2019} on how a strategic opponent can exploit them in general games, \citet{Kolumbus2022} on auctions between learning agents, and \citet{StructuralAI} on structural estimation under no-regret restrictions. Relative to that literature, which typically fixes a learning protocol or an information structure, I take the set-valued approach of robust mechanism design and bound revenue over the entire BCCE set.

\section{Model}\label{sec: model}

\subsection{Setup}
Consider a single-unit auction with a set of risk-neutral buyers $\mathcal{N} = \{1, \dots, N\}$, where $N\ge 2$. The seller owns one indivisible good and values it at zero. The buyers have a pure common value for the good, denoted $v\in V = [0,\bar{v}]\subseteq\mathbb{R}_+$, distributed according to a probability measure $\mu \in \Delta(V)$, the prior, with mean $m \coloneqq \E_\mu[v]$ and cumulative distribution function (cdf) $F$, where $\Delta(X)$ denotes the set of probability measures on $X$. Section \ref{sec: extension} treats symmetric and asymmetric priors, under which values may differ across buyers.

\begin{assumption}\label{ass:standing}
$\mu$ admits a density $f$ that is continuous and positive on $[0,\bar v]$, and each buyer's action set is $A_i = [0,\bar v]$, so that $A=\prod_i A_i = [0,\bar v]^N$.
\end{assumption}

I impose the density assumption to simplify the exposition.\footnote{Restricting bids to $[0,\bar v]$, the other part of Assumption \ref{ass:standing}, is without loss in the first-price auction, since a bid above $\bar v$ makes a loss whenever it wins. It is stated explicitly because Section \ref{sec: RGRanking} compares formats through the set of allowed deviations.} Among the results, only Proposition \ref{prop:rate}, Theorem \ref{thm:reserve} and Proposition \ref{prop:optimal_reserve} rely on it, the first through the continuity and positivity of $f$ at $0$. The action sets are kept throughout, including for priors without a density. Under Assumption \ref{ass:standing}, $F$ is continuous and strictly increasing on $[0,\bar v]$, and $F^{-1}$ denotes its inverse. Section \ref{sec: binary restriction} extends the characterization to arbitrary priors, including atomic ones. The extension is needed there because, among distributions with a given mean, the prior with the lowest guarantee is binary.

In the first-price auction (FPA), the highest bidder wins. If there are ties, the good is allocated with equal probability among the tied bidders. Formally, for a bid profile $a=(a_1,\dots,a_N)\in A$, the allocation rule $q_i(a)$ for buyer $i$ is
\begin{equation*}
    q_i(a) =
    \begin{cases}
        1, & \text{if } a_i > \max_{j \neq i} a_j, \\
        \frac{1}{| \{ k \in \mathcal{N} : a_k = \max_j a_j \} |}, & \text{if } a_i = \max_{j} a_j, \\
        0, & \text{otherwise,}
    \end{cases}
\end{equation*}
and the winner pays their own bid. Buyer $i$'s ex-post utility given the common value $v$ and the action profile $a$ is
\begin{equation*}
    u_i(v, a) = q_i(a)(v - a_i).
\end{equation*}

Since the good is always allocated, the buyers' payoffs add up to the value minus the price:
\begin{equation}\label{eq:accounting}
\textstyle\sum_{i} u_i(v,a) \;=\; v - \max_i a_i \qquad \text{for every } (v,a).
\end{equation}
Anything that lowers revenue therefore raises the buyers' total surplus by the same amount.

\subsection{Information and Two Notions of Obedience}
The buyers' correlated information about the common value is described by an \emph{information structure} $I = (S, \theta)$, where $S_i$ is a measurable space of signals for buyer $i$, $S=\prod_i S_i$, and $\theta \in \Delta(V \times S)$ is the joint distribution of values and signals. Let $\mathcal{I}(\mu)$ be the set of information structures whose value marginal is the prior, $\marg_V \theta = \mu$. The seller does not know $I$, and the guarantee is computed over all $I\in\mathcal{I}(\mu)$.

Given $I$, a \emph{decision rule} is a measurable map $b: S\to\Delta(A)$. Equivalently, a mediator observes the signal profile $s$ and privately recommends a bid $a_i$ to each buyer $i$. A decision rule induces the joint distribution
\[
\sigma_b(dv,ds,da) \;=\; b(da\mid s)\,\theta(dv,ds) \;\in\;\Delta(V\times S\times A).
\]

The two solution concepts I compare differ in one respect: \emph{how much a buyer is allowed to know when choosing a deviation}. I state both in turn.

\begin{definition}[Bayes Correlated Equilibrium]\label{def:bce}
A decision rule $b$ is a \emph{Bayes correlated equilibrium} of $I$ if for every buyer $i$ and every measurable map $\xi_i: S_i\times A_i \to A_i$,
\[
\int u_i(v,a)\,\sigma_b(dv,ds,da)\ \geq\ \int u_i\big(v,\xi_i(s_i,a_i),a_{-i}\big)\,\sigma_b(dv,ds,da).
\]
Write $\mathcal{BCE}(I)$ for the set of BCE of $I$.
\end{definition}

Definition \ref{def:bce} is the standard obedience formulation of \citet{BM2016}. The map $\xi_i$ may choose a different deviation for each signal $s_i$ and each recommended bid $a_i$, so the recommendation must be optimal conditional on both.

\begin{definition}[Bayes Coarse Correlated Equilibrium]\label{def:bcce}
A decision rule $b$ is a \emph{Bayes coarse correlated equilibrium} of $I$ if for every buyer $i$ and every \emph{constant} deviation $a_i'\in A_i$,
\[
\int u_i(v,a)\,\sigma_b(dv,ds,da)\ \geq\ \int u_i\big(v,a_i',a_{-i}\big)\,\sigma_b(dv,ds,da).
\]
Write $\mathcal{BCCE}(I)$ for the set of BCCE of $I$.
\end{definition}

In a BCCE, by contrast, the buyer considers only fixed-action deviations. No-regret learning dynamics with regret measured against fixed bids converge to this set rather than to the smaller set of BCE.\footnote{The two definitions are the extreme cases of a family ordered by how finely a deviation may condition on information. Contextual algorithms, which take value-relevant features as input, use part of that information and lead to a solution concept between BCE and BCCE. The guarantee they deliver therefore lies between the interim and the coarse guarantees, and the coarse revenue guarantee is a valid lower bound for it.\label{fn:context}} In the language used throughout the paper, Definition \ref{def:bce} imposes \emph{interim} obedience, which is evaluated after the buyer sees their information and therefore conditions on the signal. The coarse obedience of Definition \ref{def:bcce} is an \emph{ex ante} requirement, averaged over signals.

Because a constant map is a particular deviation map $\xi_i$, we have $\mathcal{BCE}(I)\subseteq\mathcal{BCCE}(I)$ for every $I$, so BCCE is the more permissive concept.\footnote{Between the two lies the notion of \citet{Hartline2015} and \citet{StructuralAI}, who impose the coarse constraint separately for each type of a player. In the present model a type is a signal, so a deviation may then depend on the buyer's own signal. This difference does not change any result of the paper. An outcome that satisfies the constraint signal by signal satisfies Definition \ref{def:bcce}, which averages the same constraint over signals. Conversely, any BCCE outcome can also be generated by an information structure in which each buyer's own signal is independent of the value and of the bids, and under such a structure the two definitions impose the same constraints. For instance, the signals can be pieces of the quantile of $v$ that are each uniform and add up to it modulo one, so the mediator, who sees all of them, can make the bids depend on $v$ while each buyer's own signal reveals nothing. Both definitions therefore generate the same joint distributions of values and bids across information structures, and the same coarse revenue guarantee. The argument relies on the freedom to choose the signals. If a deviation may condition on fixed inputs that are informative about the value, as for the contextual algorithms in footnote \ref{fn:context}, the guarantee can be higher.\label{fn:typewise}} The following example illustrates the difference.

\begin{example}[A BCCE that is not a BCE]\label{ex:gap}
Let $N=2$ and let $v$ take the values $0$ and $1$ with equal probability. Suppose both buyers observe $v$ and the mediator recommends the common bid $c=v/3$, so that both bid $0$ when $v=0$ and both bid $1/3$ when $v=1$. Each buyer's expected payoff is $\tfrac12 \cdot \tfrac12 \cdot (1-\tfrac13) = 1/6$. A buyer deviating to the fixed bid $\delta \in (0,1/3)$ wins only when $v=0$, earning $\tfrac12(0-\delta)<0$. A buyer moving to $\delta \in (1/3,1]$ always wins and earns $\tfrac12(0-\delta)+\tfrac12(1-\delta)=\tfrac12-\delta<\tfrac16$. A buyer deviating to exactly $1/3$ wins when $v=0$ and ties when $v=1$, earning $\tfrac12(0-\tfrac13)+\tfrac12\cdot\tfrac12\cdot\tfrac23=0$. Hence no fixed deviation earns more than $1/6$, and the profile is a BCCE. It is not a BCE, however. Conditional on the recommendation $c=1/3$, the buyer knows that $v=1$, and for small $\varepsilon>0$ a deviation to $1/3+\varepsilon$ wins the good and earns more than following the recommendation.
\end{example}

Example \ref{ex:gap} illustrates the mechanism behind the paper's results. Under coarse rationality, revenue can be lowered by equilibria in which bidders would gain from deviating in the states where the good is valuable but are deterred by the losses they would incur in the states where it is not. The low bid can be sustained because the coarse constraint averages the deviation payoff across states, an averaging that interim obedience does not allow.

With the two equilibrium notions in place, let $R(I,b)\coloneqq\int \max_i a_i\,\sigma_b(dv,ds,da)$ be expected revenue. The \emph{revenue guarantee} and the \emph{coarse revenue guarantee} under the prior $\mu$ are
\[
RG(\mu) \coloneqq \inf_{I\in\mathcal{I}(\mu)} \inf_{b \in \mathcal{BCE}(I)} R(I, b),
\qquad
CRG(\mu) \coloneqq \inf_{I\in\mathcal{I}(\mu)} \inf_{b \in \mathcal{BCCE}(I)} R(I, b),
\]
and since minimizing over a larger set yields a weakly smaller value, $CRG(\mu)\le RG(\mu)$ for every prior. I write $RG_N(\mu)$ and $CRG_N(\mu)$ when the number of buyers matters. I treat each infimum as the choice of an \emph{adversary}, who selects the information structure and the equilibrium to minimize revenue, and call an equilibrium that attains it a \emph{worst case}.

\subsection{A Reduced Notation for the Coarse Problem}
Definition \ref{def:bcce} has a property that Definition \ref{def:bce} lacks. Since a coarse deviation does not condition on the signal, the signals enter the constraint only through the joint distribution they induce over values and actions. The information structure can therefore be replaced, without loss, by a joint distribution $\sigma \in \Delta(V \times A)$ whose value marginal is $\mu$.

\begin{lemma}[Reduced formulation]\label{lemma:reduction_notation}
A distribution $\sigma\in\Delta(V\times A)$ with $\marg_V\sigma=\mu$ is the value-action marginal of some $\sigma_b$ with $I\in\mathcal{I}(\mu)$ and $b\in\mathcal{BCCE}(I)$ if and only if, for every buyer $i$ and every $a_i'\in A_i$,
\begin{equation}\label{eq:bcce_reduced}
\int_{V \times A} u_i(v, a) \, d\sigma(v, a) \;\geq\; \int_{V \times A} u_i(v, a'_i, a_{-i}) \, d\sigma(v, a).
\end{equation}
\end{lemma}

\begin{proof}
Necessity holds because both sides of Definition \ref{def:bcce} depend on $\sigma_b$ only through its $V\times A$ marginal. For sufficiency, let each buyer's signal be the bid that $\sigma$ recommends to them: take $S_i=A_i$, let $\theta$ be $\sigma$ itself, viewed as a law on $V\times S$, and let $b(s)$ be the point mass at $s$. Then $I\in\mathcal{I}(\mu)$, the induced law of $(v,a)$ is $\sigma$, and the BCCE constraint of Definition \ref{def:bcce} is \eqref{eq:bcce_reduced}.
\end{proof}

Write $\mathcal{BCCE}(\mu)$ for the set of $\sigma\in\Delta(V\times A)$ with $\marg_V\sigma=\mu$ that satisfy \eqref{eq:bcce_reduced}, and $R(\sigma)=\int \max_i a_i\,d\sigma$. The coarse revenue guarantee is then the infimum of revenue over this set,\footnote{I do not assume that a BCCE attaining this infimum exists. Because the payoff functions $u_i$ are discontinuous at ties, the map $\sigma\mapsto\int u_i\,d\sigma$ is not continuous in the weak-$*$ topology, and $\mathcal{BCCE}(\mu)$ need not be weak-$*$ closed, so the existence of a minimizer does not follow from compactness. Every argument below is therefore constructive. The identical play reduction maps each BCCE to an identical play BCCE with weakly lower revenue, and Theorem \ref{thm:continuous_char} then exhibits a minimizer.}
\[
CRG(\mu) \;=\; \inf_{\sigma \in \mathcal{BCCE}(\mu)} R(\sigma).
\]

\section{The Identical Play Reduction}\label{sec: identical play}

Minimizing revenue directly over $\mathcal{BCCE}(\mu)$, a set of distributions over $N$-dimensional bid profiles, may not be tractable. This section shows that it is without loss of generality to restrict the minimization to a one-dimensional family. Call $\sigma$ an \emph{identical play} distribution if it puts probability one on profiles in which all buyers submit the same bid, $a_1=\dots=a_N$.

\begin{proposition}[Identical Play Reduction]\label{prop:identical_plays}
For every $\sigma \in \mathcal{BCCE}(\mu)$ there is an identical play $\gamma \in \mathcal{BCCE}(\mu)$ with $R(\gamma)\le R(\sigma)$, and the inequality is strict unless $\sigma$ already places probability one on profiles whose two highest bids coincide. Consequently
\[
CRG(\mu) \;=\; \inf\big\{R(\gamma): \gamma\in\mathcal{BCCE}(\mu),\ \gamma \text{ identical play}\big\}.
\]
\end{proposition}

The formal argument is in Appendix \ref{app:identical}. Away from ties,\footnote{Ties need separate care, because the payoff functions jump there. If the other buyers bid $x$ with positive probability, a deviation slightly above $x$ wins in all of those states, while a deviation to $x$ itself only shares the good, so the payoff at $x$ is not the limit of nearby payoffs. Under identical play this causes no difficulty. A buyer who deviates to $x$ ties with all $N-1$ others and wins with probability $1/N$, so the payoff at $x$ lies between the two one-sided limits, and it is enough to check the coarse constraint at bids that are not submitted with positive probability. The limit from above also gives a closed version of the constraint, in which a deviation to $\delta$ wins whenever the highest competing bid is at most $\delta$, and the lower bounds use this version. Supplementary Appendix \ref{app:ties} states and proves both facts.} a buyer's payoff from a fixed deviation depends on the other bids only through the highest of them, and in any bid profile the top bidder faces the second-highest bid while every other buyer faces the highest. The \emph{pooling} construction has every buyer bid the highest competing bid of a buyer drawn at random, that is, the highest bid with probability $(N-1)/N$ and the second-highest with probability $1/N$. Each buyer then faces the highest competing bid that a randomly chosen buyer faced before, so each fixed deviation earns the buyers' average payoff from that deviation in the original equilibrium. Revenue falls, strictly unless the two highest bids coincide with probability one, so the payoff from obeying is at least the original average and the obedience constraints still hold.

With this reduction in hand, the remaining problem is to choose the joint distribution of the value $v$ and a single common bid $c$.

\section{Characterization}\label{sec: general characterization}

\subsection{The Representative-Bidder Program}\label{sec:program}

Fix an identical play distribution $\gamma\in\mathcal{BCCE}(\mu)$, and write $c$ for the common bid and $R=\E[c]$ for expected revenue. Since all $N$ bids coincide, every buyer wins with probability $1/N$ and pays $c$ when they do, so each buyer's equilibrium payoff is
\begin{equation}\label{eq:eqm_payoff}
K \;\coloneqq\; \frac{m-R}{N},
\end{equation}
the per-buyer share of the surplus not collected by the seller. I refer to $K$ as the buyers' \emph{rent}. A buyer who instead commits to the fixed bid $\delta$ wins whenever $c<\delta$ and shares when $c=\delta$, so coarse obedience requires, for every $\delta\in[0,\bar v]$,
\begin{equation}\label{eq:reduced_constraint}
K \;\ge\; \E\Big[(v-\delta)\Big(\one\{c<\delta\}+\tfrac1N\one\{c=\delta\}\Big)\Big],
\end{equation}
where $\one\{\cdot\}$ denotes the indicator function. Setting $\delta=0$ shows $K\ge0$, so revenue can never exceed the expected value.

Before stating the main result, I note two properties of the equilibrium constraint.

First, lowering revenue relaxes the constraints. By \eqref{eq:eqm_payoff}, lowering $R$ raises $K$, which appears on the left-hand side of every constraint \eqref{eq:reduced_constraint}. The guarantee is then related to a $K$ that makes the equilibrium constraint binding.

Second, a small deviation wins only in states where everyone else bids zero. Since each buyer earns $K$ by following the recommendation, such a deviation is unprofitable as long as it earns less than $K$. Buyers can therefore all bid zero at low values, where they receive the good at a price of zero, and positive bids begin only above a value threshold. The following definition names the value at which the bidding region begins.

\begin{definition}[Threshold and worst-case bidding rule]\label{def:threshold}
Let $\hat v$ denote the \emph{pooling threshold}: in the worst-case equilibrium of Theorem \ref{thm:continuous_char} below, buyers bid zero on the \emph{pooling region} $\{v<\hat v\}$ and bid a strictly increasing amount on the \emph{bidding region} $\{v\ge\hat v\}$. Write $H(v)$ for the common bid submitted at value $v\ge \hat v$.
\end{definition}

\subsection{Main Result}
\begin{theorem}\label{thm:continuous_char}
Let Assumption \ref{ass:standing} hold. The coarse revenue guarantee $CRG$ and the threshold $\hat{v}$ are the unique solution of the system
\begin{align}
\frac{m - CRG}{N} &= \int_{0}^{\hat{v}} v f(v)\,dv, \label{eq:obedience} \\
CRG &= \int_{0}^{\bar{v}} H(v) f(v)\,dv, \label{eq:revenue}
\end{align}
where $H(v)=0$ for $v<\hat v$ and, for $v\ge\hat v$,
\begin{equation}\label{eq:H}
H(v) \;=\; \frac{1}{F(v)}\int_{\hat v}^{v} z f(z)\,dz \;=\; \E\big[z \mid \hat{v} \leq z \leq v\big]\cdot \frac{F(v) - F(\hat{v})}{F(v)} ,
\end{equation}
with $z$ distributed according to $\mu$. The guarantee is strictly positive and can be attained: the identical play distribution with common bid $c=H(v)$ is a BCCE achieving it.
\end{theorem}

Appendix \ref{app:theorem1} contains the proof, which has two steps. The first step, the \emph{lower bound}, shows that no identical play BCCE yields revenue below the value given by the system. By Proposition \ref{prop:identical_plays}, every BCCE can be replaced by an identical play BCCE with no more revenue, so this value is a lower bound on the revenue of every BCCE. The bound comes from combining the obedience constraint with a rearrangement inequality, rather than from assuming that constraints bind, and I refer to the resulting inequality, \eqref{eq:chain} in Appendix \ref{app:theorem1}, as the \emph{chain}. This step also uses the \emph{uniqueness} of the solution to \eqref{eq:obedience} and \eqref{eq:revenue}, which is not immediate, since both sides of the system move in the same direction as $\hat v$ rises. It follows from a concavity argument given in the appendix. The second step, \emph{attainment}, shows that some BCCE reaches the bound. The identical play distribution with common bid $H(v)$ deters every fixed deviation, including deviations above the top of the bid range, so it is a BCCE, and its revenue is the value given by the system. That value is therefore the coarse revenue guarantee.

\paragraph{The fixed-point form}
I now write the system in an equivalent form, in the quantile variable $q=F(v)$. This form involves the prior only through its quantile function $F^{-1}(t)=\inf\{x\in[0,\bar v]:F(x)\ge t\}$, which is the inverse of $F$ under Assumption \ref{ass:standing} but exists for every prior, so the form does not need a density. For that reason Appendix \ref{app:theorem1} proves Theorem \ref{thm:continuous_char} in this form, and the same proof covers the atomic priors of Section \ref{sec: binary restriction}. Let
\begin{equation}\label{eq:Phi}
\Phi_\mu(q)\;\coloneqq\;\int_0^q F^{-1}(t)\,dt\;=\;\int_0^{F^{-1}(q)} z f(z)\,dz,
\qquad q\in[0,1],
\end{equation}
the expected value collected on the lowest $q$ quantiles of the prior, where the second equality uses the density. It is convex and nondecreasing, with $\Phi_\mu(0)=0$ and $\Phi_\mu(1)=m$. Define the revenue functional
\begin{equation}\label{eq:Rcal}
\mathcal{R}_\mu(K)\;\coloneqq\;\int_0^1 \frac{\big(\Phi_\mu(q)-K\big)^+}{q}\,dq, \qquad K\in[0,m],
\end{equation}
where $x^+\coloneqq\max\{x,0\}$. In this notation, the rearrangement inequality says that $\E[v\one\{B\}]\ge\Phi_\mu(q)$ for every event $B$ of probability $q$, and the chain says that in an identical play BCCE with rent $K$ the quantile function $C$ of the common bid satisfies $C(q)\ge(\Phi_\mu(q)-K)/q$ for every $q\in(0,1)$. Integrating the chain gives $R\ge\mathcal{R}_\mu(K)$. The system of \eqref{eq:obedience} and \eqref{eq:revenue} is equivalent to the single equation
\begin{equation}\label{eq:fixed_point}
m-NK \;=\; \mathcal{R}_\mu(K)
\end{equation}
in the rent: the unique solution is $K^*=\int_0^{\hat v}vf(v)\,dv$, written $K^*_N$ when the number of buyers matters, the guarantee is $CRG=m-NK^*=\mathcal{R}_\mu(K^*)$, and the worst-case bid is $c=\psi(F(v))=H(v)$, where
\begin{equation}\label{eq:psi}
\psi(q)\;\coloneqq\;\frac{\big(\Phi_\mu(q)-K^*\big)^+}{q}.
\end{equation}
The functional $\mathcal{R}_\mu$ is also used in the bounds of Section \ref{sec: extension} and in the convex-order comparison of Section \ref{sec: binary restriction}.

Equation \eqref{eq:obedience} is the obedience constraint for a bid just above zero. Its right-hand side is the expected value collected on the pooling region, which is also the payoff from deviating to such a bid. The rent $K$ must be large enough to make that deviation unattractive, and in the worst case it is exactly large enough.

Equation \eqref{eq:H} describes the bidding rule above the threshold. The common bid at value $v$ is the expected value conditional on lying in the bidding region and below $v$, shaded by the probability that a win comes from the bidding region rather than from the pooling region. Comparing it with the limiting BCE bidding rule $G(v)=\E[z\mid z\le v]$ shows the source of the gap:
\[
G(v)-H(v) \;=\; \frac{K^*}{F(v)} \qquad \text{for } v\ge\hat v .
\]
Above the threshold, the coarse bidding rule is the limiting interim rule minus the term $K^*/F(v)$. The gap between $CRG$ and the limiting guarantee of Theorem \ref{thm:asymptotic} is therefore the integral of this term against $f$ over the bidding region, plus the integral of $G$ against $f$ over the pooling region.

\begin{example}[Uniform values]\label{ex:uniform}
Let $\mu=U[0,1]$ be the uniform distribution on $[0,1]$, so $m=1/2$ and $F(v)=v$. Then $H(v)=(v^2-\hat v^2)/(2v)$, and the system reduces to the single equation
\[
\tfrac12-\tfrac{N\hat v^2}{2} \;=\; \tfrac14-\tfrac{\hat v^2}{4}+\tfrac{\hat v^2\log \hat v}{2}.
\]
With two buyers the threshold is $\hat v\approx0.6730$ and $CRG\approx0.0471$, so the seller collects less than $10\%$ of the good's expected value. Since $F(\hat v)=\hat v$ for the uniform distribution, the pooling region has probability $0.673$. In the worst case both buyers bid zero on this region, and the good is sold at a price of zero. Conditional on a positive price being paid, the average price is $0.144$. With ten buyers, $CRG\approx0.1916$ and the pooling probability falls to $0.248$.
\end{example}

\begin{proposition}[Comparative statics in $N$]\label{prop:monotoneN}
If $\mu$ is non-degenerate, then $CRG_N(\mu)$ is strictly increasing in $N$, and the worst-case rent $K^*_N$ is strictly decreasing in $N$.
\end{proposition}

\begin{proof}
In \eqref{eq:fixed_point}, the left side $m-NK$ is the revenue that remains when each buyer earns the rent $K$, and the right side $\mathcal{R}_\mu(K)$ is the lower bound that the chain places on revenue at that rent, which does not depend on $N$. By Lemma \ref{lemma:concavity} in Appendix \ref{app:theorem1}, $g_N(K)\coloneqq m-NK-\mathcal{R}_\mu(K)$ is positive below $K^*_N$ and negative above it. An additional buyer lowers the left side by $K$, so $g_{N+1}(K^*_N)=-K^*_N<0$, and therefore $K^*_{N+1}<K^*_N$. Since $\mathcal{R}_\mu$ is strictly decreasing, $CRG_{N+1}(\mu)=\mathcal{R}_\mu(K^*_{N+1})>\mathcal{R}_\mu(K^*_N)=CRG_N(\mu)$.
\end{proof}

The guarantee increases in $N$ for a reason that seems specific to coarse rationality. The rent that deters a fixed deviation is a per-buyer quantity: each buyer receives $K$, and by \eqref{eq:eqm_payoff} the total is $NK=m-R$. With more buyers, each one's share of a given amount of uncollected surplus is smaller, so a given level of revenue leaves each buyer with a smaller equilibrium payoff, and the constraints permit a smaller revenue reduction.

\subsection{Asymptotic Convergence and Its Rate}

\begin{theorem}\label{thm:asymptotic}
Under Assumption \ref{ass:standing},
\[
\lim_{N \to \infty} CRG_N(\mu) \;=\; \int_{0}^{\bar{v}} G(v) f(v)\,dv \;=\; \lim_{N \to \infty} RG_N(\mu),
\qquad G(v) = \E[z \mid z \leq v].
\]
\end{theorem}

The proof is in Appendix \ref{app:asymptotic}. Intuitively, as $N$ grows, the worst-case rent $K^*_N$ vanishes at rate $1/N$, the pooling threshold $\hat v$ falls to zero, and by the identity $G-H=K^*_N/F$ the coarse bidding rule converges pointwise to the limiting interim rule $G$ characterized by \citet{BBM2017}.

The next question is the rate of convergence. For pure common values, \citet{BBM2017} characterize the minimum-revenue BCE. In the worst case the winning bid is a deterministic function of the value, and the bid and the revenue are
\begin{equation}\label{eq:bbm_finite}
\beta_N(v)\;=\;\frac{\int_0^v z\,dF^{\frac{N-1}{N}}(z)}{F^{\frac{N-1}{N}}(v)},
\qquad
RG_N(\mu)\;=\;\int_0^{\bar v}\beta_N(v)\,f(v)\,dv,
\end{equation}
where $\beta_N(v)$ is the conditional mean of the value under the tilted distribution $F^{(N-1)/N}$.\footnote{In the as-if reading, each of $N$ independent values has marginal $F^{1/N}$, so that the highest of them is distributed as $F$ and the highest rival value as $F^{(N-1)/N}$. Then $\beta_N$ is the symmetric equilibrium bid of that model, the expected highest rival value conditional on winning.} For the uniform prior, $\beta_N(v)=\frac{N-1}{2N-1}\,v$ and $RG_N=\frac{N-1}{2(2N-1)}$, which equals $1/6$ at $N=2$ and increases to $1/4$, so in this example the gap to the limit is of order $1/N$. The next result shows that the interim guarantee converges at rate $1/N$ for every prior, and that the coarse guarantee converges only at the slower rate $\log N/N$.

\begin{proposition}[Rate of convergence]\label{prop:rate}
Let Assumption \ref{ass:standing} hold and let $R_\infty$ denote the limit in Theorem \ref{thm:asymptotic}. Then $0\le R_\infty-RG_N\le 2\bar v/(N-1)$, while
\[
R_\infty - CRG_N \;=\; \frac{(m-R_\infty)\,\log N}{2N}\,\big(1+o(1)\big).
\]
Consequently, the cost of coarseness satisfies
\[
RG_N - CRG_N \;=\; \frac{(m-R_\infty)\,\log N}{2N}\,\big(1+o(1)\big):
\]
the interim guarantee converges at rate $1/N$, the coarse guarantee at rate $\log N/N$, and the gap between them closes at $\log N/N$.
\end{proposition}

The proof is in Supplementary Appendix \ref{app:rate}. Write $q_N\coloneqq F(\hat v)$ for the pooling quantile with $N$ buyers, so that $\Phi_\mu(q_N)=K^*_N$. The gap $R_\infty-CRG_N$ equals the rent $K^*_N$ times $\log(1/q_N)$, up to a term of order $1/N$, and $K^*_N$ is of order $1/N$ for every prior. The rate therefore depends on how fast $q_N$ goes to zero, and this is the one place where the continuity and positivity of $f$ at $0$ in Assumption \ref{ass:standing} are used. Near zero, $\Phi_\mu(q)\approx q^2/(2f(0))$, hence $q_N\approx\sqrt{2f(0)K^*_N}$, and the square root produces the factor $\tfrac12$ in the rate.

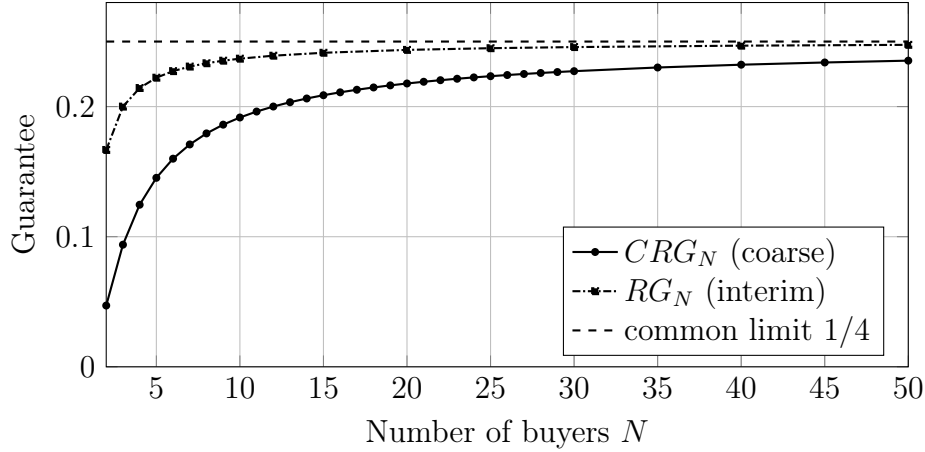
\begin{figure}[H]
\centering
\begin{tikzpicture}
\begin{axis}[width=0.8\textwidth, height=0.42\textwidth,
  xlabel={Number of buyers $N$}, ylabel={Guarantee},
  xmin=2, xmax=50, ymin=0, ymax=0.28,
  legend pos=south east, legend cell align=left, grid=major]
\addplot[thick, mark=*, mark size=1.1pt] coordinates {
(2,0.04709) (3,0.09387) (4,0.12457) (5,0.14526) (6,0.15997) (7,0.17092) (8,0.17939) (9,0.18613) (10,0.19163) (11,0.19620) (12,0.20006) (13,0.20337) (14,0.20624) (15,0.20875) (16,0.21098) (17,0.21295) (18,0.21472) (19,0.21632) (20,0.21777) (21,0.21909) (22,0.22030) (23,0.22141) (24,0.22244) (25,0.22338) (26,0.22427) (27,0.22508) (28,0.22585) (29,0.22657) (30,0.22724) (35,0.23006) (40,0.23221) (45,0.23392) (50,0.23532)};
\addlegendentry{$CRG_N$ (coarse)}
\addplot[thick, densely dashdotted, mark=square*, mark size=1.1pt] coordinates {
(2,0.16667) (3,0.20000) (4,0.21429) (5,0.22222) (6,0.22727) (7,0.23077) (8,0.23333) (9,0.23529) (10,0.23684) (12,0.23913) (15,0.24138) (20,0.24359) (25,0.24490) (30,0.24576) (40,0.24684) (50,0.24747)};
\addlegendentry{$RG_N$ (interim)}
\addplot[dashed, thick, domain=2:50] {0.25};
\addlegendentry{common limit $1/4$}
\end{axis}
\end{tikzpicture}
\caption{Interim and coarse revenue guarantees for $\mu=U[0,1]$}
\label{fig:convergence}
\end{figure}

Table \ref{tab:uniform} and Figure \ref{fig:convergence} illustrate these rates for the uniform prior. With two bidders the coarse guarantee is under thirty percent of the interim guarantee, and with ten it is about eighty percent.

\begin{table}[H]
\centering
\small
\setlength{\tabcolsep}{3pt}
\begin{tabular}{lccccccc}
\toprule
$N$ & 2 & 3 & 5 & 10 & 20 & 50 & 100 \\
\midrule
Interim guarantee $RG_N$ & 0.1667 & 0.2000 & 0.2222 & 0.2368 & 0.2436 & 0.2475 & 0.2487 \\
Coarse guarantee $CRG_N$ & 0.0471 & 0.0939 & 0.1453 & 0.1916 & 0.2178 & 0.2353 & 0.2419 \\
Cost of coarseness $RG_N-CRG_N$ & 0.1196 & 0.1061 & 0.0770 & 0.0452 & 0.0258 & 0.0122 & 0.0068 \\
$CRG_N$ below $RG_N$ (\%) & 71.7 & 53.1 & 34.6 & 19.1 & 10.6 & 4.9 & 2.7 \\
Pooling probability $F(\hat v)$ & 0.673 & 0.520 & 0.377 & 0.248 & 0.168 & 0.103 & 0.072 \\
\bottomrule
\end{tabular}
\caption{Interim and coarse revenue guarantees for $\mu=U[0,1]$, both converging to $1/4$.}
\label{tab:uniform}
\end{table}

The strict inequality between $RG_N$ and $CRG_N$ is not specific to the uniform prior. The next result shows that it holds for every prior and every finite $N$, provided the interim minimum is attained as in \citet{BBM2017}.

\begin{proposition}[Strict coarseness gap]\label{prop:strict}
Let Assumption \ref{ass:standing} hold and suppose the infimum defining $RG_N(\mu)$ is attained by a BCE in which the two highest bids differ with positive probability, as in the construction of \citet{BBM2017}. Then
\[
CRG_N(\mu)\;<\;RG_N(\mu) .
\]
\end{proposition}

\begin{proof}
Let $\sigma_B$ be the value-action marginal of a BCE attaining $RG_N(\mu)$ whose two highest bids differ with positive probability. In the construction of \citeauthor{BBM2017} this holds because the winning bid is a deterministic function of the value and the losing bids lie strictly below it. Since $\mathcal{BCE}(I)\subseteq\mathcal{BCCE}(I)$ for every $I$, Lemma \ref{lemma:reduction_notation} gives $\sigma_B\in\mathcal{BCCE}(\mu)$, with revenue $RG_N(\mu)$, and Proposition \ref{prop:identical_plays} turns it into an identical play BCCE with strictly lower revenue. Hence $CRG_N(\mu)<RG_N(\mu)$.
\end{proof}

\section{Reserve Prices}\label{sec: design}

When $N$ is small, the worst case of Sections \ref{sec: identical play} and \ref{sec: general characterization} sells the good at a price of zero with high probability. This section shows that a reserve price can raise the guarantee substantially.

Formally, fix a reserve price $r\in[0,\bar v)$. The first-price auction with reserve $r$ allocates the good to the highest bidder at the bid price if the highest bid is at least $r$, breaking ties uniformly, and withholds the good otherwise. Write $CRG_N(r)$ for the resulting coarse revenue guarantee.

Bids below $r$ never win, so without loss of generality every bid is either $0$, interpreted as staying out, or at least $r$. Since the good is no longer always allocated, equation \eqref{eq:accounting} becomes
\begin{equation}\label{eq:accounting_r}
\textstyle\sum_i u_i(v,a) \;=\; (v-a_{(1)})\,\one\{a_{(1)}\ge r\},
\end{equation}
where $a_{(1)}\ge a_{(2)}\ge\dots\ge a_{(N)}$ are the order statistics of the bid profile.

The pooling construction of Proposition \ref{prop:identical_plays} cannot be applied directly. Suppose exactly one buyer clears the reserve, so $a_{(1)}\ge r>a_{(2)}$. Pooling onto $a_{(2)}$ with probability $1/N$ would then withhold the good, and by \eqref{eq:accounting_r} this removes the buyers' surplus together with the seller's revenue, which can tighten the obedience constraints. A modified construction still works. It pools onto $\max\{a_{(2)},r\}$ instead, so that the good is still sold, and switches to staying out only where the pooled bid equals the reserve and exceeds the value.

\begin{proposition}[Identical Play with a Reserve]\label{prop:reduction_r}
Fix $r\in[0,\bar v)$. For every BCCE $\sigma$ of the first-price auction with reserve $r$ there is an identical play BCCE $\gamma$ with $R(\gamma)\le R(\sigma)$, where the common bid takes values in $\{0\}\cup[r,\bar v]$.
\end{proposition}

The proof is in Appendix \ref{app:reserve}.

Under a reserve, buyers can no longer receive the good at a price of zero, but they have a less costly alternative: pooling at exactly $r$, where the good still trades and the winner pays only the reserve. For values below $r$, however, buyers prefer not to buy at all. The worst case therefore involves up to two pools: an \emph{opt-out pool} at zero, where the good is not sold, and a \emph{reserve pool} at exactly $r$, where the good trades at the reserve. The opt-out pool occupies the lowest $s$ quantiles, the reserve pool the next $q_0-s$, and above $q_0$ the common bid is strictly increasing. Call the identical play distributions of this form the \emph{two-pool family}, and let $K$ again denote each buyer's equilibrium payoff, the rent. As in \eqref{eq:psi}, the bid at a quantile $q>q_0$ is $(\Phi_\mu(q)-K)/q$. The family is determined as follows. A bid just above the reserve wins against both pools outright, so obedience requires $K\ge\Phi_\mu(q_0)-rq_0$. The bidding rule must also start at or above the reserve, $(\Phi_\mu(q_0)-K)/q_0\ge r$, which is the same inequality reversed. Hence, in the worst case,
\begin{equation}\label{eq:capture}
K \;=\; \Phi_\mu(q_0)-r\,q_0 .
\end{equation}
The opt-out pool removes the value $\Phi_\mu(s)$ from the buyers' total surplus, so
\begin{equation}\label{eq:family}
NK \;=\; m-\Phi_\mu(s)-R,
\qquad
R \;=\; r\,(q_0-s)+\int_{q_0}^{1}\frac{\Phi_\mu(q)-K}{q}\,dq .
\end{equation}
Given the opt-out share $s$ and the rent $K$, \eqref{eq:capture} places $q_0$ where the increasing part of the bidding rule crosses the reserve, so the bid at every quantile $q>s$ is $\max\{r,(\Phi_\mu(q)-K)/q\}$. Equations \eqref{eq:capture} and \eqref{eq:family} therefore reduce to a single equation in the rent,
\begin{equation}\label{eq:kappa}
NK \;=\; m-\Phi_\mu(s)-\int_s^1\max\Big\{r,\frac{\Phi_\mu(q)-K}{q}\Big\}\,dq .
\end{equation}
By \eqref{eq:family}, revenue is $m-\Phi_\mu(s)-NK$, so for a given opt-out share the worst case uses the largest solution $K=\kappa_+(s)$ of \eqref{eq:kappa} in $[0,m]$. Under the condition of Theorem \ref{thm:reserve} below, $\kappa_+(s)<m-r$, so the increasing part of the bidding rule does cross the reserve at some $q_0<1$. Call $s$ \emph{feasible} if $\kappa_+(s)$ exists and $\Phi_\mu(s)-rs\le\kappa_+(s)$, which then says that the reserve pool $[s,q_0)$ is well defined and that the average value in it is at least $r$. Let
\[
\widehat V_N(r)\;\coloneqq\;\inf\big\{\,m-\Phi_\mu(s)-N\kappa_+(s)\;:\;s\ \text{feasible}\,\big\}.
\]

\begin{theorem}[Reserve prices]\label{thm:reserve}
Let Assumption \ref{ass:standing} hold and let $r\in(0,m)$ satisfy $rF(r)-\Phi_\mu(F(r))<(N-1)(m-r)$, which is true for every $r\le m(N-1)/N$. Then
\[
CRG_N(r)\;=\;\widehat V_N(r),
\]
so the two-pool family is the worst case. For $r\ge m$ the guarantee is zero, since no participation is a BCCE.
\end{theorem}

The proof is in Appendix \ref{app:reserve}. The lower bound follows the chain of Theorem \ref{thm:continuous_char} with two changes. The quantile function of the common bid is now bounded below by $\max\{r,(\Phi_\mu(q)-K)/q\}$ rather than by $(\Phi_\mu(q)-K)/q$, and equation \eqref{eq:accounting_r} removes the value of the states in which buyers opt out, which by the rearrangement inequality is at least $\Phi_\mu(s)$. For a given opt-out share, the largest rent consistent with both bounds defines a member of the two-pool family, which attains the bound.

\begin{example}[Uniform values with a reserve]\label{ex:uniform_r}
Return to the setting of Example \ref{ex:uniform}, where $\mu=U[0,1]$, $N=2$, and without a reserve $CRG_2\approx0.047$, sustained by both buyers bidding zero on about the lowest two-thirds of values and receiving the good at a price of zero. Table \ref{tab:profile} traces the worst case as the reserve grows. At $r=0.01$ buyers bid $r$ on nearly the entire former pooling region, almost no one opts out, and the guarantee rises by less than the reserve itself, from about $0.047$ to $0.055$. As the reserve grows, pooling at the reserve becomes costly for the buyers and the worst case shifts toward opting out, so the share $s$ rises. The guarantee is maximized at $r\approx0.148$, the reserve at which the reserve pool vanishes.\footnote{Theorem \ref{thm:reserve} covers every $r<\sqrt2-1\approx0.414$ in this example. For $\sqrt2-1\le r<1/2$, solving \eqref{eq:capture} and \eqref{eq:family} with $s=q_0$ gives a member of the two-pool family without a reserve pool whose rent is below $m-r$, so the argument of Step 2 of the proof of Theorem \ref{thm:reserve} shows that it is a BCCE, and its revenue is below $0.05$. For $r\ge1/2$ the guarantee is zero.} Below that level, part of the pooling region still bids $r$. A larger reserve lowers the probability of sale and does not tighten obedience further. At this reserve the guarantee is $CRG_2\approx0.106$, more than double the guarantee without a reserve. The gain accrues gradually, and reaching half of it requires $r\approx0.041$.
\end{example}

\begin{table}[H]
\centering
\begin{tabular}{lccccc}
\toprule
$r$ & 0.01 & 0.05 & 0.10 & 0.148 & 0.20 \\
\midrule
$CRG_2(r)$ & 0.0552 & 0.0812 & 0.1000 & 0.1056 & 0.1038 \\
opt-out share $s$ & 0.051 & 0.253 & 0.464 & 0.621 & 0.664 \\
pool boundary $q_0$ & 0.676 & 0.674 & 0.650 & 0.621 & 0.664 \\
\bottomrule
\end{tabular}
\caption{The worst case across reserve levels, $\mu=U[0,1]$, $N=2$.}
\label{tab:profile}
\end{table}

Example \ref{ex:uniform_r} suggests that the guarantee is largest at a reserve where the reserve pool vanishes, and the next result shows that this is a necessary condition for an optimal reserve in the range of Theorem \ref{thm:reserve}.

\begin{proposition}[Optimal reserve]\label{prop:optimal_reserve}
Let Assumption \ref{ass:standing} hold, and let $r^*\in(0,m)$ maximize $CRG_N(r)$ over $r\in[0,\bar v)$ and satisfy the condition of Theorem \ref{thm:reserve}. Then $CRG_N(r^*)$ is attained by a member of the two-pool family without a reserve pool, $s=q_0$, in which buyers with values below $F^{-1}(s)$ opt out and the others bid above $r^*$.
\end{proposition}

The proof is in Appendix \ref{app:reserve}. A higher reserve has two effects on the worst case. It raises the price paid in the reserve pool, and it makes bidding the reserve a less attractive deviation, so that more buyers can be made to opt out. While the worst case has a reserve pool, the adversary could let more buyers opt out but prefers to pool them at the reserve, so only the first effect matters at the margin. For a given opt-out share $s$, a higher reserve lowers the rent $\kappa_+(s)$ at the rate $(q_0-s)/(N-\log(1/q_0))$, so revenue $m-\Phi_\mu(s)-N\kappa_+(s)$ rises at $N$ times that rate.

When the reserve pool is empty, no buyer pays exactly the reserve, and only the second effect matters. The adversary then lets as many buyers opt out as the reserve allows, and a higher reserve, which allows more, can lower the guarantee. The proof uses only the direction of the first effect. If every worst case at a reserve had a reserve pool, a slightly higher reserve would raise the guarantee, so at an optimal reserve the guarantee is attained without one.\footnote{Having no reserve pool is necessary but not sufficient, since in Example \ref{ex:uniform_r} the worst case has none at every reserve above $r^*$ in the range of Theorem \ref{thm:reserve}. For some priors the reserve pool also vanishes, reappears and vanishes again as the reserve rises, so the smallest reserve at which it vanishes need not be optimal.} In Example \ref{ex:uniform_r} the optimal reserve is $r^*\approx0.148$, where the reserve pool vanishes and the opt-out share is about $0.621$.

\section{Beyond Common Values: Symmetric and Asymmetric Priors}\label{sec: extension}

The analysis so far, with and without a reserve, assumes pure common values. The interim characterization of \citet{BBM2017} covers symmetric priors with arbitrary correlation, and this section asks how much of the coarse analysis extends to values that may differ across buyers. Let $\pi$ be a prior on $V^N$ that is exchangeable, that is, invariant under every permutation of the coordinates, with common per-buyer marginal $\mu$. I also call such priors symmetric. Buyer $i$'s payoff is $q_i(a)(v_i-a_i)$, and a BCCE is defined as in Lemma \ref{lemma:reduction_notation}, now over joint laws of $(v_1,\dots,v_N,a_1,\dots,a_N)$ whose marginal on the value profile is $\pi$. Such a law is exchangeable if it is invariant under permuting values and bids together, $(v,a)\mapsto(v_\rho,a_\rho)$ with $(v_\rho)_i=v_{\rho(i)}$ and $(a_\rho)_i=a_{\rho(i)}$, for every bijection $\rho$ of $\{1,\dots,N\}$. A joint law of $(v_1,\dots,v_N)$ with given marginals is called a coupling. Pure common values are the coupling $v_1=\dots=v_N$, and independent private values are the opposite extreme. Write $CRG(\pi)$ for the infimum of revenue over all BCCE under $\pi$. Under common values, Proposition \ref{prop:identical_plays} reduces this infimum to identical play, but the pooling argument behind it does not extend when values differ.\footnote{The contrast with the interim characterization, which handles every symmetric coupling, comes from how each solution concept constrains a bidder. In a BCE, obedience is checked conditional on the bidder's own recommendation, so the adversary can correlate each bidder's signal with that bidder's own value. This gives the adversary $N$ separate instruments, one per bidder, which is why the worst case of \citet{BBM2017} handles arbitrary symmetric couplings and is even allocatively efficient. A bidder's own bid appears nowhere in the coarse deviation payoff $\E[(v_i-\delta)\one\{\max_{j\ne i}a_j<\delta\}]$, so under BCCE buyer $i$ is constrained only through the correlation between $v_i$ and the other buyers' bids.} This section explains why and gives the lower bound that remains valid, first for symmetric priors and then for asymmetric ones.

Averaging a BCCE over relabelings of the buyers gives an exchangeable BCCE with the same revenue, because the prior is exchangeable and the coarse constraints are linear, so the obstacle is the pooling step. The pooling construction of Proposition \ref{prop:identical_plays} allocates the good to a uniformly random buyer instead of to the top bidder. Under common values this changes nothing, since every buyer values the good equally. When values differ, it changes two things. First, if the top bidder has an above-average value, reassigning the good at random lowers the buyers' total surplus, so the pooled rent can fall rather than rise. Second, before pooling, a deviation that outbids everyone except the top bidder wins only if the deviator is the top bidder, while after pooling it wins for every buyer with probability $1/N$, so the deviation payoff can change. When the top bidder's value is either never above the average or always the highest, the two corrections have the same sign and the first is at least as large in absolute value. Pooling therefore still works when the symmetrized distribution allocates against value but can fail when it allocates efficiently. With heterogeneous values the adversary can lower revenue by allocating efficiently: the good goes to the buyer with the highest value, the buyers' total surplus rises, and the additional surplus relaxes the obedience constraints, which permits lower bids.

Moreover, in the first-price auction the adversary can do this at an arbitrarily small cost in revenue. Fix a common price and give the buyer with the highest value a bid premium of $\varepsilon$ above it. The premium decides the allocation and adds only $\varepsilon$ to revenue, while every coarse deviation still faces, up to $\varepsilon$, the common price as its highest competing bid. The lower bound that remains valid follows the chain of Theorem \ref{thm:continuous_char}, with the mean value replaced by the expected highest value. Let $m_N\coloneqq\E_\pi[\max_iv_i]$ denote the expected highest value, so that $m\le m_N\le Nm$, and let $\tilde K$ be the unique solution in $(0,m]$ of
\begin{equation}\label{eq:fixed_point_pi}
m_N-NK\;=\;\mathcal{R}_\mu(K),
\end{equation}
which exists by the argument of Lemma \ref{lemma:concavity} in Appendix \ref{app:theorem1} with $m_N$ in place of $m$.

\begin{theorem}[Symmetric priors]\label{thm:symmetric}
Let $\pi$ be exchangeable with per-buyer marginal $\mu$ satisfying Assumption \ref{ass:standing}. Then
\[
CRG(\pi)\;\ge\;m_N-N\tilde K\;=\;\mathcal{R}_\mu(\tilde K).
\]
\end{theorem}

The proof is in Appendix \ref{app:ipv}. The bound follows the chain of Theorem \ref{thm:continuous_char} with two substitutions. The rearrangement inequality is applied to the pair $(v_i,M_i)$ of a buyer's own value and highest competing bid, which uses only the marginal $\mu$, and the analog of \eqref{eq:accounting} bounds the buyers' total surplus by the efficient surplus $m_N-R$ rather than by $m-R$. When $m_N>m$, identical play does not attain the bound. Under identical play the good goes to a random buyer, so each buyer's rent is $(m-R)/N$, and the argument of Theorem \ref{thm:continuous_char} gives revenue at least $\mathcal{R}_\mu(K^*)$. This exceeds the bound $\mathcal{R}_\mu(\tilde K)$, because at $K^*$ the left side of \eqref{eq:fixed_point_pi} exceeds the right side by $m_N-m>0$, so $\tilde K>K^*$. Whether some other BCCE attains the bound is left open.

Since the proof uses the prior only through the marginal $\mu$ and through $m_N$, the theorem holds for every prior with common per-buyer marginal $\mu$, exchangeable or not. When the marginals differ, the same two steps give a bound in which each buyer has their own chain. Let $\pi$ be any prior on $V^N$ with per-buyer marginals $\mu_1,\dots,\mu_N$, write $m_{\mu_i}\coloneqq\E[v_i]$, define $\Phi_{\mu_i}$ and $\mathcal{R}_{\mu_i}$ by \eqref{eq:Phi} and \eqref{eq:Rcal} with $\mu_i$ in place of $\mu$, and let
\begin{equation}\label{eq:Rinv}
\mathcal{R}_{\mu_i}^{-1}(R)\;\coloneqq\;\inf\big\{K\in[0,m_{\mu_i}]:\mathcal{R}_{\mu_i}(K)\le R\big\},
\end{equation}
the smallest rent at which buyer $i$'s chain permits revenue $R$. Since $\mathcal{R}_{\mu_i}$ is continuous and strictly decreasing from $\mathcal{R}_{\mu_i}(0)$ to $0$ on $[0,m_{\mu_i}]$, this is the inverse of $\mathcal{R}_{\mu_i}$ for $R\le\mathcal{R}_{\mu_i}(0)$ and equals zero above.

\begin{theorem}[Asymmetric priors]\label{thm:asymmetric}
Let $\pi$ be a prior on $V^N$ whose per-buyer marginals $\mu_1,\dots,\mu_N$ each satisfy Assumption \ref{ass:standing}, and let $m_N\coloneqq\E_\pi[\max_iv_i]$. Then $CRG(\pi)\ge\tilde R$, where $\tilde R$ is the smallest $R\in[0,m_N]$ satisfying
\begin{equation}\label{eq:asym_bound}
\sum_{i=1}^N\mathcal{R}_{\mu_i}^{-1}(R)\;\le\;m_N-R .
\end{equation}
When $\mu_i=\mu$ for every $i$, $\tilde R=m_N-N\tilde K$, so Theorem \ref{thm:symmetric} is the case of identical marginals.
\end{theorem}

The proof is in Appendix \ref{app:ipv}. Each buyer's chain says that revenue $R$ requires buyer $i$'s rent to be at least $\mathcal{R}_{\mu_i}^{-1}(R)$, and the surplus bound says that the rents sum to at most $m_N-R$, so $\tilde R$ is the smallest revenue at which the rents that the chains require fit within the surplus. With different marginals the adversary can distribute the total rent unevenly, giving more to the buyers whose chains are harder to satisfy.

\section{Ranking and Equivalence Across Auctions}\label{sec: RGRanking}

I now return to common values and compare the first-price auction with other formats, following \citet{BBM2019RGE}. As in their framework, a \emph{standard auction} is one that allocates the good to the highest bidder and admits an equilibrium in monotonic pure strategies in the symmetric independent private values model. Ties are broken uniformly, losers pay nothing, and the winner pays $p(a)$, which I take to satisfy $0\le p(a)\le a_{(1)}$, with equality whenever $a_{(1)}=a_{(2)}$: the winner never pays more than their bid, and a winner who is tied with another bidder pays the tied bid. Let $\mathcal{M}$ denote this class, and write $CRG(M)$ for the coarse revenue guarantee of $M\in\mathcal{M}$. Within it, the pricing rules that matter are defined by what happens on profiles in which the $N-1$ losing bids coincide. A standard auction has \emph{own-bid pricing} if the winner pays $a_{(1)}$ whenever the other bids coincide, as in the first-price and Dutch auctions, and \emph{competing-bid pricing} if the winner pays $a_{(2)}$ whenever the other bids coincide, as in the second-price and English auctions.

The comparison combines two facts. First, identical play is an outcome that every standard auction can produce at the same revenue. In \citeauthor{BBM2019RGE} this role is played by an information structure under which every standard auction has an equilibrium whose revenue is the first-price guarantee. Second, formats differ only in what a single deviator pays after outbidding the others.

\begin{proposition}[Identical play is format-free]\label{prop:identical_equivalence}
Fix any identical play distribution with common bid $c$. In every standard auction the good is allocated uniformly among the $N$ buyers, the winner pays $c$, and expected revenue is $\E[c]$. In every standard auction with own-bid pricing, moreover, the identical play BCCE are exactly those of the first-price auction, so the coarse revenue guarantee restricted to identical play equals $CRG(\mathrm{FPA})$ in each of them.
\end{proposition}

\begin{proof}
Under own-bid pricing, a buyer who deviates from an identical play distribution to the fixed bid $\delta$ loses if $\delta<c$, ties at price $c$ if $\delta=c$, and wins at price $\delta$ if $\delta>c$, exactly as in the first-price auction, and equilibrium payoffs coincide by the first part. Hence the coarse constraints \eqref{eq:bcce_reduced} are the same in both formats, and the sets of identical play BCCE coincide.
\end{proof}

\begin{proposition}[Coarse revenue guarantee ranking]\label{prop:ranking}
The first-price auction has the highest coarse revenue guarantee among standard auctions with own-bid or competing-bid pricing. For every $M\in\mathcal{M}$ with own-bid pricing, $CRG(M)\le CRG(\mathrm{FPA})$, and for every $M\in\mathcal{M}$ with competing-bid pricing, in particular the second-price and English auctions,
\[
CRG(M)\;=\;0\;<\;CRG(\mathrm{FPA}).
\]
\end{proposition}

\begin{proof}
Under own-bid pricing, the identical play BCCE attaining $CRG(\mathrm{FPA})$ in Theorem \ref{thm:continuous_char} is a BCCE of $M$ with the same revenue, by Proposition \ref{prop:identical_equivalence}, so $CRG(M)\le CRG(\mathrm{FPA})$. Under competing-bid pricing, consider the distribution in which one designated buyer bids $\bar v$ and all others bid zero, independently of the value. Revenue is zero. The designated buyer earns $m$ and cannot improve, since any positive bid still wins at the price $p(\delta,0,\dots,0)=0$. Any other buyer must bid $\bar v$ to contest the good, which ties the top bid and, since a winner who is tied pays the tied bid, means winning with some probability $\lambda\in(0,1]$ at price $\bar v$, for a payoff of $\lambda(m-\bar v)\le0$, and the conclusion does not depend on how ties are broken. This is a BCCE with zero revenue, so $CRG(M)=0$, and $CRG(\mathrm{FPA})>0$ by Theorem \ref{thm:continuous_char}.
\end{proof}

The zero-revenue equilibrium in this proof is an instance of the ``bidding ring'' equilibria discussed by \citet{BBM2019RGE}. It is a Bayes Nash equilibrium under every information structure, so the interim guarantee of these formats is zero as well.

\section{Unknown Value Distribution}\label{sec: binary restriction}

So far the seller has been assumed to know the prior. Suppose instead that the seller knows only the expected value $E=\E[v]$ and the upper bound $\bar v$ on the value, but not the shape of the distribution. The worst-case prior is then binary, which violates Assumption \ref{ass:standing}, so I first extend the characterization to priors without a density.

\begin{theorem}[Beyond the density]\label{thm:general}
Let $\mu$ be any non-degenerate distribution on $[0,\bar v]$, that is, one not concentrated on a single value, and read the definitions \eqref{eq:Phi}, \eqref{eq:Rcal}, and \eqref{eq:psi} with the quantile function $F^{-1}(t)=\inf\{x\in[0,\bar v]: F(x)\ge t\}$ in place of the inverse. Then \eqref{eq:fixed_point} has a unique solution $K^*\in(0,m)$, and $CRG(\mu)=m-NK^*=\mathcal{R}_\mu(K^*)>0$. The infimum is attained by the identical play distribution $v=F^{-1}(U)$, $c=\psi(U)$, with $U$ uniform on $[0,1]$.
\end{theorem}

Appendix \ref{app:theorem1} proves Theorem \ref{thm:general} together with Theorem \ref{thm:continuous_char}. In the attaining distribution, the uniform variable is the rank of the value. Away from the atoms of $\mu$ we have $U=F(v)$, so the coupling reduces to the deterministic bid $H(v)=\psi(F(v))$ of Theorem \ref{thm:continuous_char}. Conditional on an atom of $\mu$, however, $U$ is uniform on the corresponding quantile interval, so the common bid is randomized continuously across it. The randomization is required by the constraints. Minimality requires the obedience constraint to bind at every deviation level, and a deviation to $\psi(q)$ with $q$ inside an atom's interval must win exactly a fraction of the atom, which a single common bid cannot arrange. If the common bid were the same across the whole atom, a deviation just above it would win the entire atom at once and earn strictly more than the rent.

With the characterization extended to atomic priors, the relevant comparison across priors is the convex order. Write $\mu\cx\nu$ if $\int\phi\,d\mu\le\int\phi\,d\nu$ for every convex function $\phi$ on $[0,\bar v]$, or equivalently if $\mu$ and $\nu$ have the same mean and $\nu$ is a mean-preserving spread of $\mu$. I read $\mu\cx\nu$ as ``$\nu$ is riskier than $\mu$.''

\begin{proposition}[Convex order monotonicity]\label{prop:convex_order}
If $\mu\cx\nu$ then $CRG(\nu)\le CRG(\mu)$, for every $N$. Consequently, for every $N$,
\[
\min_{\mu\in\Delta([0,\bar v]):\ \E_\mu[v]=E} CRG(\mu) \;=\; CRG(\underline\mu),
\]
where $\underline\mu$ is the binary distribution with $\underline\mu(\{0\})=1-E/\bar v$ and $\underline\mu(\{\bar v\})=E/\bar v$.
\end{proposition}

The proof is in Supplementary Appendix \ref{app:binary}. A low price is sustained in the worst case because the bid increases with the value, so a buyer who deviates to a slightly higher bid wins mostly in states where the value is low, and the deviation is unprofitable despite the low price. A mean-preserving spread lowers the expected value on the lowest quantiles, $\Phi_\nu\le\Phi_\mu$, which makes such deviations even less profitable. Lower bids then satisfy the obedience constraints, and revenue falls. Since the binary distribution on $\{0,\bar v\}$ is the riskiest distribution on $[0,\bar v]$ with a given mean, it is the seller's worst case.

\begin{corollary}[The binary guarantee]\label{cor:binary}
Let $p_0\coloneqq\underline\mu(\{0\})=1-E/\bar v$ be the probability that $\underline\mu$ assigns to the value zero. For finite $N$, $CRG(\underline\mu)=E-N\bar v(\hat q-p_0)$, where $\hat q\in(p_0,1)$ is the unique solution of $(N-1)(p_0-q)=q\log q$. As $N\to\infty$,
\begin{equation}\label{eq:binary_limit}
CRG(\underline\mu) \;\longrightarrow\; E + (\bar{v}-E)\log\Big(1-\frac{E}{\bar{v}}\Big).
\end{equation}
\end{corollary}

\section{Conclusion}\label{sec: conclusion}

This paper characterizes the worst-case revenue of auctions when bidders are rational only in the coarse sense that no single fixed bid would have earned them more on average. No-regret learning leads to this restriction, and it seems well suited to algorithmic bidders. In the first-price auction a worst case is an identical play equilibrium, in which bidders do not use their private information and all submit the same bid, and the reduction to identical play yields a closed-form characterization of the guarantee. The guarantee is strictly positive, increasing in the number of bidders, and strictly below the interim benchmark of \citet{BBM2017} at every finite market size. Both converge to a common limit, and the coarse guarantee does so more slowly, at rate $\log N/N$ rather than $1/N$. When values differ across buyers, the pooling argument behind the identical play reduction does not extend, but the coarse revenue guarantee is still bounded below by the solution of a system of equations similar to the one for common values, with one chain of constraints per buyer.

The worst case is sustained by states in which every bidder bids zero and the buyers receive the good at a price of zero, and the resulting surplus is what satisfies their obedience constraints. A reserve price reduces this surplus, and coarse bidders respond by pooling at the reserve and pay the reserve price when it is worth doing so. At a reserve that maximizes the guarantee, the guarantee is attained by a worst case without a reserve pool, in which every buyer either opts out or bids above the reserve. Across the formats studied, the first-price auction has the highest guarantee. Its worst case is available at the same revenue in every standard auction and remains an equilibrium whenever a deviator pays their own bid, and when a deviator pays the others' bid instead, the guarantee of that format is zero. Finally, the guarantee is monotone in the convex order, so the binary distribution is the worst case among priors with a given mean and upper bound.

For a seller who faces learning algorithms, the results suggest three things. The first-price auction is a natural choice, since it has the highest guarantee among the formats studied. A reserve price is most useful at a level at which a worst case involves opting out but no pooling at the reserve. And when the value distribution is unknown, the binary distribution with the same mean is the appropriate benchmark.

Several questions still remain. Away from common values, the paper gives only a lower bound on the guarantee, and it is not known whether the bound is attained, or by which coupling of bids and values. Whether the first-price auction maximizes the coarse revenue guarantee over all standard auctions, as it maximizes the interim guarantee in \citet{BBM2019RGE}, is also open, since the ranking proved here covers only the formats in which a deviator who outbids tied opponents pays either their own bid or the opponents' bid. The optimal mechanism against coarsely rational bidders has not been characterized either, and the reserve price studied here is only one simple instrument. Finally, the guarantee for contextual algorithms lies between the coarse and the interim guarantees (footnote \ref{fn:context}), and characterizing it is left for future work.

\newpage

\appendix

\section{Proofs for Section \ref{sec: identical play}}\label{app:identical}

\subsection*{Proof of Proposition \ref{prop:identical_plays}}

Throughout, $a_{(1)}\ge a_{(2)}\ge\dots\ge a_{(N)}$ are the order statistics of a bid profile $a$, and $M_i\coloneqq\max_{j\neq i}a_j$ is buyer $i$'s highest competing bid. The proposition follows from a single pooling step, in which every buyer bids the highest competing bid of a buyer drawn at random. Under identical play every buyer faces the same coarse constraint, so it is enough that the constraints of $\sigma$ hold on average across buyers.

\begin{lemma}[Pooling]\label{lemma:pooling}
Let $\sigma\in\mathcal{BCCE}(\mu)$, let $J$ be uniform on $\{1,\dots,N\}$ and independent of $(v,a)$, and let $\gamma$ be the law of $(v,M_J,\dots,M_J)$. Then $\gamma\in\mathcal{BCCE}(\mu)$ is identical play, and
\[
R(\gamma)=\E_{\sigma}\Big[\tfrac{N-1}{N}a_{(1)}+\tfrac1N a_{(2)}\Big]\;\le\;R(\sigma),
\]
strictly unless $a_{(1)}=a_{(2)}$ holds $\sigma$-almost surely.
\end{lemma}
\begin{proof}
By definition, $\gamma$ is identical play with value marginal $\mu$. Write $c\coloneqq M_J$. Given $(v,a)$, the top bidder, if unique, faces $a_{(2)}$ and every other buyer faces $a_{(1)}$, so $c$ equals $a_{(1)}$ with probability $\frac{N-1}{N}$ and $a_{(2)}$ otherwise, which remains true under a tie. This gives the formula for $R(\gamma)$, and $c\le a_{(1)}$ gives $R(\gamma)\le R(\sigma)$, with equality only if $a_{(1)}=a_{(2)}$ almost surely.

Fix $\delta$ that is not an atom of the law of $c$, which is the average of the laws of the $M_i$. Under $\sigma$ a deviation to $\delta$ then ties with probability zero, so buyer $i$'s constraint reads $\E_\sigma[(v-\delta)\one\{M_i<\delta\}]\le\E_\sigma[u_i]$, and averaging over buyers gives
\[
\E_\gamma\big[(v-\delta)\one\{c<\delta\}\big]\;=\;\frac1N\sum_{i}\E_{\sigma}\big[(v-\delta)\one\{M_i<\delta\}\big]\;\le\;\frac1N\sum_{i}\E_{\sigma}[u_i] .
\]
By \eqref{eq:accounting}, the buyers' total payoff is $v-a_{(1)}$ under $\sigma$ and $v-c\ge v-a_{(1)}$ under $\gamma$, where each buyer receives $1/N$ of it, so $K_\gamma\coloneqq\E_\gamma[u_i]\ge\frac1N\sum_i\E_\sigma[u_i]$. Hence $K_\gamma\ge\E_\gamma[(v-\delta)\one\{c<\delta\}]$ at every such $\delta$, and $\gamma\in\mathcal{BCCE}(\mu)$ by Lemma \ref{lemma:ties}.
\end{proof}

Lemma \ref{lemma:pooling} proves the proposition without assuming that a minimizer exists. \qed

\section{Proofs for Section \ref{sec: general characterization}}\label{app:theorem1}

I prove Theorem \ref{thm:continuous_char} in its fixed-point form. Lemma \ref{lemma:concavity} below shows that \eqref{eq:fixed_point} has a unique solution $K^*\in(0,m)$, and the proof then has two steps. Step 1 shows that $m-NK^*$ is a lower bound on the revenue of every identical play BCCE, and hence, by Proposition \ref{prop:identical_plays}, on the revenue of every BCCE. Step 2 shows that the identical play distribution with $v=F^{-1}(U)$ and $c=\psi(U)$, for $U$ uniform on $[0,1]$, is a BCCE whose revenue equals this bound, so $m-NK^*$ is the coarse revenue guarantee. Under Assumption \ref{ass:standing}, $U=F(v)$ almost surely, so this is the coupling $c=\psi(F(v))$, which is \emph{comonotone} in the sense that the common bid is a nondecreasing function of the value. The substitution $q=F(v)$, under which $\Phi_\mu(F(\hat v))=K^*$ and $\psi(F(v))=H(v)$, then turns these claims into the theorem. The density is used only for this translation, and neither Proposition \ref{prop:identical_plays} nor Lemma \ref{lemma:ties} uses it, so the proof also establishes Theorem \ref{thm:general}. In this appendix, therefore, $\mu$ is any non-degenerate prior on $[0,\bar v]$ and $F^{-1}(t)=\inf\{x\in[0,\bar v]:F(x)\ge t\}$ is its quantile function.

The function $\Phi_\mu$ is convex, nondecreasing, and continuous, with $\Phi_\mu(0)=0$ and $\Phi_\mu(1)=m$. For $K\in[0,m]$, the pooling quantile $q_K\coloneqq\sup\{q\in[0,1]:\Phi_\mu(q)\le K\}$ satisfies $\Phi_\mu(q_K)=K$, and $\Phi_\mu(q)>K$ exactly for $q>q_K$. Since $\Phi_\mu(p)/p$ is the average of the nondecreasing function $F^{-1}$ over $[0,p]$, for every $K\ge0$ the map
\begin{equation}\label{eq:slope}
p\;\mapsto\;\frac{\Phi_\mu(p)-K}{p}\;=\;\frac{\Phi_\mu(p)}{p}-\frac{K}{p}
\end{equation}
is continuous and nondecreasing on $(0,1]$. It is strictly increasing if $K>0$, and also if $K=0$ under Assumption \ref{ass:standing}, since $F^{-1}$ is then strictly increasing.

\begin{lemma}[Rearrangement bound]\label{lemma:HL}
For any event $B$ with $\Pr(B)=q$, $\E[v\one\{B\}]\ge\Phi_\mu(q)$.
\end{lemma}
\begin{proof}
This is the Hardy--Littlewood rearrangement inequality \citep{HLP1934}. Among events of probability $q$, the expectation $\E[v\one\{B\}]$ is smallest when $B$ is placed on the lowest realizations of $v$, splitting an atom of $\mu$ if necessary, which gives $\int_0^q F^{-1}(t)dt$.
\end{proof}

\begin{lemma}[Concavity and uniqueness]\label{lemma:concavity}
$\mathcal{R}_\mu$ is continuous, convex, and strictly decreasing on $[0,m]$, with $\mathcal{R}_\mu(m)=0$. Hence $g_N(K)\coloneqq m-NK-\mathcal{R}_\mu(K)$ is concave on $[0,m]$ and has a unique zero $K^*\in(0,m)$, with $g_N>0$ on $[0,K^*)$ and $g_N<0$ on $(K^*,m]$.
\end{lemma}
\begin{proof}
Each integrand $K\mapsto(\Phi_\mu(q)-K)^+/q$ is convex and nonincreasing, strictly decreasing near $K$ for every $q\in(q_K,1]$, a set of positive length when $K<m$, and bounded by $\Phi_\mu(q)/q\le F^{-1}(q)$. Hence $\mathcal{R}_\mu$ is convex, strictly decreasing, and, by dominated convergence, continuous, and $\mathcal{R}_\mu(m)=0$ because $\Phi_\mu\le m$. Thus $g_N$ is concave, with $g_N(m)=-(N-1)m<0$ and $g_N(0)=m-\int_0^1\Phi_\mu(q)/q\,dq>0$, since the average $\Phi_\mu(q)/q$ of $F^{-1}$ over $[0,q]$ is strictly below its average $m$ over $[0,1]$ for $q<1$ when $\mu$ is non-degenerate. A concave function that is positive at $0$ and negative at $m$ has a unique zero and changes sign there.
\end{proof}

\paragraph{Step 1: a lower bound for every BCCE.} Obedience against each fixed deviation forces the common bid to be high enough at every quantile, given the rent, and integrating these requirements bounds revenue from below. Fix an identical play $\gamma\in\mathcal{BCCE}(\mu)$ with common bid $c$, revenue $R=\E[c]$, and rent $K=(m-R)/N\ge0$, and let $F_c$ and $C(q)\coloneqq\inf\{x\in[0,\bar v]:F_c(x)\ge q\}$ be the cdf and the quantile function of $c$. By Lemma \ref{lemma:ties} without a reserve and with $M_i=c$, the closed constraint $K\ge\E[(v-\delta)\one\{c\le\delta\}]$ holds for every $\delta\in[0,\bar v]$, and by Lemma \ref{lemma:HL} its right side is at least $\Phi_\mu(F_c(\delta))-\delta F_c(\delta)$. Fix $q\in(0,1)$ and take $\delta=C(q)$, so that $p\coloneqq F_c(C(q))\ge q$ by right-continuity. Then $K\ge\Phi_\mu(p)-C(q)\,p$, and since $p\ge q$, \eqref{eq:slope} gives
\begin{equation}\label{eq:chain}
C(q)\;\ge\;\frac{\Phi_\mu(p)-K}{p}\;\ge\;\frac{\Phi_\mu(q)-K}{q}.
\end{equation}
Since also $C\ge0$, integrating gives $R=\int_0^1C(q)\,dq\ge\mathcal{R}_\mu(K)$. With $R=m-NK$, this says $g_N(K)\ge0$. By Lemma \ref{lemma:concavity}, $K\le K^*$, so $R\ge m-NK^*$. Every identical play BCCE therefore yields revenue at least $m-NK^*$, and since Proposition \ref{prop:identical_plays} replaces any BCCE by an identical play BCCE with no more revenue, so does every BCCE.

\paragraph{Step 2: a BCCE attains the bound.} It remains to find a BCCE whose revenue is exactly $m-NK^*$. Let $U$ be uniform on $[0,1]$, $v=F^{-1}(U)$, and $c=\psi(U)$. Revenue is $\int_0^1\psi=\mathcal{R}_\mu(K^*)=m-NK^*$, so the rent is $K^*$. Since $\Phi_\mu(q_{K^*})=K^*$, the bid $\psi$ is $0$ on $[0,q_{K^*}]$, and by \eqref{eq:slope} it is continuous and strictly increasing on $(q_{K^*},1]$, rising from $0$ to $\psi(1)=m-K^*$. A deviation to $\delta\in(0,m-K^*]$ therefore wins exactly when $U<q$, where $\psi(q)=\delta$, and ties with probability zero, so it earns $\E[(F^{-1}(U)-\delta)\one\{U<q\}]=\Phi_\mu(q)-\delta q=K^*$. A deviation to $\delta>m-K^*$ always wins and earns $m-\delta<K^*$. A deviation to $0$ ties with the pool and earns $\frac1N\E[v\one\{c=0\}]=\Phi_\mu(q_{K^*})/N=K^*/N<K^*$. Hence the coupling is a BCCE with revenue $m-NK^*$. Together with Step 1, this gives $CRG(\mu)=m-NK^*=\mathcal{R}_\mu(K^*)$, which is positive because $K^*<m$. \qed

\subsection*{Asymptotic convergence}\label{app:asymptotic}

\begin{proof}[Proof of Theorem \ref{thm:asymptotic}]
From $CRG_N=m-NK^*_N\ge0$ we get $K^*_N\le m/N\to0$. The integrand in \eqref{eq:Rcal} satisfies $0\le\frac{(\Phi_\mu(q)-K)^+}{q}\le\frac{\Phi_\mu(q)}{q}\le\bar v$ and converges pointwise to $\frac{\Phi_\mu(q)}{q}$ as $K\downarrow0$. Dominated convergence gives $CRG_N=\mathcal{R}_\mu(K^*_N)\to\int_0^1\frac{\Phi_\mu(q)}{q}dq$, and substituting $q=F(v)$ turns this into $\int_0^{\bar v}G(v)f(v)dv$, the limiting revenue guarantee of \citet{BBM2017}.
\end{proof}

\section{Reserve Prices}\label{app:reserve}

\subsection*{Proof of Proposition \ref{prop:reduction_r}}

For $r=0$ this is Proposition \ref{prop:identical_plays}, so let $r>0$. A bid in $[0,r)$ never wins, so setting it to $0$ changes no payoff, and I take every bid to lie in $\{0\}\cup[r,\bar v]$. Fix a BCCE $\sigma$, let $J$ be as in Lemma \ref{lemma:pooling}, and write $\bar K_\sigma\coloneqq\frac1N\sum_i\E_\sigma[u_i]$, which is nonnegative because a deviation to $0$ never wins. With a reserve, I modify the common bid $M_J$ of Lemma \ref{lemma:pooling} in two steps. First, on the event $\{a_{(1)}\ge r\}$ that the good is sold, I raise it to $\max\{M_J,r\}$. On the complement every bid is $0$, so $M_J=0$ and nothing changes. Second, where the result equals $r$ but the value is below $r$, I replace it by $0$, so that the buyers stay out. Let $c$ be the final common bid and $\gamma$ the identical play distribution in which every buyer bids $c$. Then $\gamma$ has value marginal $\mu$, the common bid takes values in $\{0\}\cup[r,\bar v]$, $v\ge r$ on $\{c=r\}$, and $c$ differs from $M_J$ only where both are at most $r$.

\emph{Revenue and payoffs.} On $\{a_{(1)}\ge r\}$ both $M_J$ and $r$ are at most $a_{(1)}$, so $c\le a_{(1)}$, and $c=0$ elsewhere. Hence $R(\gamma)\le R(\sigma)$. By \eqref{eq:accounting_r}, the buyers' total payoff is $(v-a_{(1)})\one\{a_{(1)}\ge r\}$ under $\sigma$ and $(v-c)\one\{c\ge r\}$ under $\gamma$. On $\{a_{(1)}\ge r\}$ the first step replaces the price $a_{(1)}$ by $\max\{M_J,r\}\le a_{(1)}$. Where the second step applies, the total changes from $v-r<0$ to $0$. Hence the total weakly rises in every state. Under $\gamma$ each buyer receives $1/N$ of the total, so $K_\gamma\ge\bar K_\sigma\ge0$.

\emph{Deviations.} A deviation to $\delta<r$ never wins and earns $0\le K_\gamma$. For $\delta>r$, the events $\{c<\delta\}$ and $\{M_J<\delta\}$ coincide, and so do $\{c=\delta\}$ and $\{M_J=\delta\}$. Hence, at every $\delta>r$ that is not an atom of the law of $c$, the argument in the proof of Lemma \ref{lemma:pooling} gives
\[
\E_\gamma\big[(v-\delta)\one\{c<\delta\}\big]\;=\;\frac1N\sum_i\E_\sigma\big[(v-\delta)\one\{M_i<\delta\}\big]\;\le\;\bar K_\sigma\;\le\;K_\gamma .
\]
By Lemma \ref{lemma:ties}, every deviation to $\delta\in(r,\bar v]$ earns at most $K_\gamma$, and $K_\gamma\ge B\coloneqq\E_\gamma[(v-r)\one\{c\le r\}]$. The lemma does not reach the deviation to exactly $r$, because bids below the reserve never win. A deviator bidding $r$ wins outright when $c=0$ and ties the pool when $c=r$, so this deviation earns $A+(B-A)/N$, where $A\coloneqq\E_\gamma[(v-r)\one\{c=0\}]$. Since $v\ge r$ on $\{c=r\}$, $B-A\ge0$, and the deviation earns at most $B\le K_\gamma$. Hence $\gamma$ is a BCCE. \qed

\subsection*{Proof of Theorem \ref{thm:reserve}}

For $r\ge m$, no participation is a BCCE, since a deviation to any $\delta\ge r$ wins and earns $m-\delta\le0$. Fix now $r\in(0,m)$.

\emph{Notation.} The function $q\mapsto\Phi_\mu(q)-rq$ vanishes at $0$, decreases on $[0,F(r)]$, and increases thereafter, so for $\kappa\ge0$ the set $\{q:\Phi_\mu(q)-rq\le\kappa\}$ is an interval $[0,q_r(\kappa)]$ with $q_r(\kappa)\ge F(r)$, and $(\Phi_\mu(q)-\kappa)/q>r$ exactly for $q>q_r(\kappa)$. Let $\underline C_\kappa(q)\coloneqq\max\{r,(\Phi_\mu(q)-\kappa)/q\}$, which is nonincreasing in $\kappa$, and let $h_s(\kappa)\coloneqq m-\Phi_\mu(s)-N\kappa-\int_s^1\underline C_\kappa(q)\,dq$, so that \eqref{eq:kappa} reads $h_s(\kappa)=0$. The map $(s,\kappa)\mapsto h_s(\kappa)$ is continuous on $[0,1]\times[0,m]$ with $h_s(m)<0$, so whenever $h_s(\kappa)\ge0$ for some $\kappa$, the largest solution $\kappa_+(s)$ of \eqref{eq:kappa} exists and is at least $\kappa$. Moreover, $\kappa_+(s)<m-r$. For $\kappa\ge m-r$ we have $(\Phi_\mu(q)-\kappa)/q\le m-\kappa\le r$, so $\underline C_\kappa\equiv r$ and $h_s(\kappa)\le h_s(m-r)=rs-\Phi_\mu(s)-(N-1)(m-r)$, which is negative by the hypothesis of the theorem because $rs-\Phi_\mu(s)\le rF(r)-\Phi_\mu(F(r))$. Write $\mathcal F(s)\coloneqq m-\Phi_\mu(s)-N\kappa_+(s)$, which equals $\int_s^1\underline C_{\kappa_+(s)}(q)\,dq$ because $h_s(\kappa_+(s))=0$.

\emph{Step 1: the lower bound.} Fix an identical play BCCE with common bid $c\in\{0\}\cup[r,\bar v]$, revenue $R$, opt-out share $s\coloneqq\Pr(c=0)$, bid quantile function $C$, and rent $K$, which is nonnegative because a deviation to $0$ never wins. By Lemma \ref{lemma:ties}, the closed constraint $K\ge\E[(v-\delta)\one\{c\le\delta\}]$ holds for $\delta\in[r,\bar v]$, and the derivation of the chain \eqref{eq:chain} uses it only at $\delta=C(q)$. Hence $C(q)\ge(\Phi_\mu(q)-K)/q$ wherever $C(q)\ge r$, that is, for $q>s$, and $R=\int_s^1C\ge\int_s^1\underline C_K$. By \eqref{eq:accounting_r} and Lemma \ref{lemma:HL}, $NK=m-\E[v\one\{c=0\}]-R\le m-\Phi_\mu(s)-R$. Together, $h_s(K)\ge0$, so $K\le\kappa_+(s)$ and $R\ge\int_s^1\underline C_{\kappa_+(s)}=\mathcal F(s)$. At $\delta=r$, the closed constraint and Lemma \ref{lemma:HL} give $\Phi_\mu(q_1)-rq_1\le K$ for $q_1\coloneqq\Pr(c\le r)$, so $s\le q_1\le q_r(K)$ and $\Phi_\mu(s)-rs\le K\le\kappa_+(s)$. Thus $s$ is feasible, and by Proposition \ref{prop:reduction_r}, $CRG_N(r)\ge\inf\{\mathcal F(s):s\text{ feasible}\}$.

\emph{Step 2: attainment.} Let $s$ be feasible, $\kappa\coloneqq\kappa_+(s)$, and $q_0\coloneqq q_r(\kappa)$. Since $\kappa<m-r=\Phi_\mu(1)-r$, we have $q_0<1$ and, by continuity, $\Phi_\mu(q_0)-rq_0=\kappa$, which is \eqref{eq:capture}, and feasibility gives $s\le q_0$. The two-pool candidate with opt-out share $s$ is the comonotone identical play distribution with common bid $C(F(v))$, where $C=0$ on $[0,s)$ and $C=\underline C_\kappa$ on $[s,1]$, so that $C=r$ on $[s,q_0]$ and $C(q)=(\Phi_\mu(q)-\kappa)/q$ on $(q_0,1]$. Its revenue is $\mathcal F(s)$, so by \eqref{eq:accounting_r} each buyer earns $\kappa$. Since $q_0\ge F(r)>0$, \eqref{eq:capture} gives $\lim_{q\downarrow q_0}C(q)=r$, so the argument of Step 2 of Appendix \ref{app:theorem1}, applied to $C$ on $(q_0,1]$, shows that every deviation $\delta>r$ earns at most $\kappa$. A deviation below $r$ never wins. A deviation to $r$ wins on the opt-out pool and ties the reserve pool, so by \eqref{eq:capture} and feasibility it earns
\[
\Phi_\mu(s)-rs+\tfrac1N\big(\kappa-\Phi_\mu(s)+rs\big)\;=\;\kappa-\big(1-\tfrac1N\big)\big(\kappa-\Phi_\mu(s)+rs\big)\;\le\;\kappa .
\]
Hence the candidate is a BCCE with revenue $\mathcal F(s)$, and $CRG_N(r)=\inf\{\mathcal F(s):s\text{ feasible}\}$.

\emph{Step 3: the minimum is attained.} The feasible set is nonempty by Step 1, applied for instance to the BCCE in which every buyer bids $m$. If feasible shares $s_n$ converge to $s$ and $\kappa_+(s_n)$ converges to $\kappa$, continuity gives $h_s(\kappa)=0$ and $\Phi_\mu(s)-rs\le\kappa\le\kappa_+(s)$. Hence the feasible set is compact and $\mathcal F$ is lower semicontinuous on it, so $\mathcal F$ attains its minimum, and $CRG_N(r)=\widehat V_N(r)$. \qed

\subsection*{Proof of Proposition \ref{prop:optimal_reserve}}

To keep track of the reserve, write $h^r_s$, $\kappa^r_+(s)$ and $\mathcal F^r(s)$ for the objects of the proof of Theorem \ref{thm:reserve} at reserve $r$. A feasible share $s$ has a reserve pool at $r$ if $q_r(\kappa^r_+(s))>s$, and otherwise $q_0=q_r(\kappa^r_+(s))=s$.

\emph{Step 1: at a fixed opt-out share with a reserve pool, a higher reserve lowers the rent.} Let $r<r'$, and let $s$ be feasible at $r$ with a reserve pool, so that $\kappa\coloneqq\kappa^r_+(s)$ satisfies $q_r(\kappa)>s$. Since $\max\{r',x\}\ge\max\{r,x\}$, we have $h^{r'}_s\le h^r_s$. Moreover, on $(s,q_r(\kappa))$ we have $(\Phi_\mu(q)-\kappa)/q\le r$, so the integrand in $h^{r'}_s(\kappa)$ exceeds that in $h^r_s(\kappa)$ by $r'-r$ there, and $h^{r'}_s(\kappa)\le h^r_s(\kappa)-(r'-r)\big(q_r(\kappa)-s\big)<0$. Since $\kappa$ is the largest zero of $h^r_s$ and $h^r_s(m)<0$, $h^r_s<0$ on $(\kappa,m]$. Hence $h^{r'}_s<0$ on $[\kappa,m]$, so if $s$ is also feasible at $r'$, then $\kappa^{r'}_+(s)<\kappa$ and $\mathcal F^{r'}(s)>\mathcal F^r(s)$.

\emph{Step 2: conclusion.} Suppose, to the contrary, that every feasible share that minimizes $\mathcal F^{r^*}$ has a reserve pool at $r^*$. Both sides of the condition of Theorem \ref{thm:reserve} are continuous in $r$, so the condition holds on $[r^*,r^*+\varepsilon)$ for some $\varepsilon>0$. Take $r_n\downarrow r^*$ with $r_n\in(r^*,r^*+\varepsilon)$. By Theorem \ref{thm:reserve} and Step 3 of its proof, there are feasible shares $s_n$ at $r_n$ with $\mathcal F^{r_n}(s_n)=CRG_N(r_n)\le CRG_N(r^*)$, where the inequality is the optimality of $r^*$. Write $\kappa_n\coloneqq\kappa^{r_n}_+(s_n)$, and pass to a subsequence along which $s_n\to\bar s$ and $\kappa_n\to\bar\kappa$. Since $h^r_s(\kappa)$ is continuous in $(s,\kappa)$, $|h^r_s(\kappa)-h^{r'}_s(\kappa)|\le|r-r'|$, and $\Phi_\mu(s_n)-r_ns_n\le\kappa_n$ for every $n$, we get $h^{r^*}_{\bar s}(\bar\kappa)=0$ and $\Phi_\mu(\bar s)-r^*\bar s\le\bar\kappa\le\kappa^{r^*}_+(\bar s)$. Hence $\bar s$ is feasible at $r^*$, and
\[
\mathcal F^{r^*}(\bar s)\;\le\;m-\Phi_\mu(\bar s)-N\bar\kappa\;=\;\lim_{n}CRG_N(r_n)\;\le\;CRG_N(r^*)\;=\;\min\mathcal F^{r^*}.
\]
So $\bar s$ minimizes $\mathcal F^{r^*}$ and $\kappa^{r^*}_+(\bar s)=\bar\kappa$, and by hypothesis $q_{r^*}(\bar\kappa)>\bar s$. Since $h^{r^*}_{s_n}(\kappa_n)\ge h^{r_n}_{s_n}(\kappa_n)=0$, we have $\kappa^{r^*}_+(s_n)\ge\kappa_n$. The map $q_{r^*}$ is nondecreasing, and it is continuous because $\Phi_\mu(q)-r^*q$ is continuous and strictly increasing on $[F(r^*),1]$, so $q_{r^*}(\kappa^{r^*}_+(s_n))\ge q_{r^*}(\kappa_n)>s_n$ for large $n$. For such $n$, $s_n$ is feasible at $r^*$ with a reserve pool, and Step 1 gives
\[
CRG_N(r_n)\;=\;\mathcal F^{r_n}(s_n)\;>\;\mathcal F^{r^*}(s_n)\;\ge\;CRG_N(r^*),
\]
a contradiction. Hence some feasible share $s$ that minimizes $\mathcal F^{r^*}$ has $q_0=s$, and by Step 2 of the proof of Theorem \ref{thm:reserve} the corresponding member attains $CRG_N(r^*)$. In it, buyers with values below $F^{-1}(s)$ opt out, and since $(\Phi_\mu(q)-\kappa^{r^*}_+(s))/q>r^*$ exactly for $q>q_{r^*}(\kappa^{r^*}_+(s))=s$, the others bid above $r^*$. \qed

\section{Symmetric and Asymmetric Priors}\label{app:ipv}

Throughout, $\pi$ is a prior on $V^N$ with per-buyer marginals $\mu_1,\dots,\mu_N$, a BCCE is a joint law $\sigma$ of $(v_1,\dots,v_N,a_1,\dots,a_N)$ with value marginal $\pi$ satisfying the coarse constraints, $u_i=q_i(a)(v_i-a_i)$, $M_i\coloneqq\max_{j\ne i}a_j$, and $K_i\coloneqq\E_\sigma[u_i]$. Since the good is always allocated, $\sum_iu_i=v_{\mathrm{win}}-a_{(1)}$, where $v_{\mathrm{win}}\coloneqq\sum_iq_i(a)v_i$ is the expected value of the winner.

\subsection*{Proof of Theorem \ref{thm:asymmetric}}

The set of $R\in[0,m_N]$ satisfying \eqref{eq:asym_bound} is closed, since each $\mathcal{R}_{\mu_i}^{-1}$ is continuous by Lemma \ref{lemma:concavity} applied to $\mu_i$, and it is nonempty, since at $R=\max_i\mathcal{R}_{\mu_i}(0)$ every term on the left is zero while $\mathcal{R}_{\mu_i}(0)<m_{\mu_i}\le m_N$. Hence $\tilde R$ is well defined.

Fix a BCCE $\sigma$ with revenue $R=\E[a_{(1)}]$ and rents $K_1,\dots,K_N$. For each buyer $i$, Lemma \ref{lemma:ties} gives $K_i\ge\E[(v_i-\delta)\one\{M_i\le\delta\}]$ for every $\delta\in[0,\bar v]$, so $K_i\ge0$ by the case $\delta=0$. The argument for \eqref{eq:chain} and for $R\ge\mathcal{R}_\mu(K)$ in Step 1 of Appendix \ref{app:theorem1}, applied to the pair $(v_i,M_i)$ in place of $(v,c)$, uses only the marginal $\mu_i$ of $v_i$, so it gives
\begin{equation}\label{eq:chain_pi}
\E[M_i]\;\ge\;\mathcal{R}_{\mu_i}(K_i)\qquad\text{for every buyer }i.
\end{equation}
Since $M_i\le a_{(1)}$ pointwise, $R\ge\mathcal{R}_{\mu_i}(K_i)$, and since $\mathcal{R}_{\mu_i}$ is nonincreasing this forces $K_i\ge\mathcal{R}_{\mu_i}^{-1}(R)$, for otherwise $\mathcal{R}_{\mu_i}(K_i)>R$. Since the winner's value is at most the highest value, $\sum_iK_i=\E[v_{\mathrm{win}}-a_{(1)}]\le m_N-R$. Summing, $\sum_i\mathcal{R}_{\mu_i}^{-1}(R)\le\sum_iK_i\le m_N-R$, so $R$ satisfies \eqref{eq:asym_bound} and $R\ge\tilde R$. Taking the infimum over BCCE gives the inequality. The last claim of the theorem is proved with Theorem \ref{thm:symmetric} below. \qed

\subsection*{Proof of Theorem \ref{thm:symmetric}}

By Theorem \ref{thm:asymmetric} with $\mu_i=\mu$ for every $i$, it suffices to show that $\tilde R=m_N-N\tilde K$. Let $\tilde g(K)\coloneqq m_N-NK-\mathcal{R}_\mu(K)=g_N(K)+(m_N-m)$. By Lemma \ref{lemma:concavity}, $\tilde g$ is concave, with $\tilde g(0)=g_N(0)+(m_N-m)>0$ because $m_N\ge m$, and $\tilde g(m)=m_N-Nm\le0$ because $\max_iv_i\le\sum_iv_i$. Hence $\tilde g\ge0$ exactly on $[0,\tilde K]$. For $R\in[0,m_N]$ the rent $K\coloneqq(m_N-R)/N$ lies in $[0,m]$, and since $\mathcal{R}_\mu$ is continuous and nonincreasing, \eqref{eq:asym_bound} reads $\mathcal{R}_\mu^{-1}(R)\le K$, that is, $\mathcal{R}_\mu(K)\le R$, that is, $\tilde g(K)\ge0$, that is, $R\ge m_N-N\tilde K$. Hence $\tilde R=m_N-N\tilde K$. \qed

\newpage
\bibliographystyle{aea}
\bibliography{refCRGFPA}

\clearpage
\setcounter{section}{0}
\renewcommand{\thesection}{S.\arabic{section}}
\renewcommand{\theHsection}{S.\arabic{section}}
\begin{center}
{\Large\bfseries Supplementary Appendix}
\end{center}

\noindent This supplement begins with the treatment of ties used throughout the appendices. It then collects the proofs of the rate of convergence (Proposition \ref{prop:rate}), the convex-order comparison (Proposition \ref{prop:convex_order}), and the binary guarantee (Corollary \ref{cor:binary}). Notation is that of the main appendix.

\section{Ties}\label{app:ties}

\begin{lemmaS}[Ties and the closed constraint]\label{lemma:ties}
Fix a reserve $r\in[0,\bar v)$, with $r=0$ meaning no reserve, a joint law of values and bids, and a buyer $i$ with value $v_i$, expected payoff $K_i$, and highest competing bid $M_i$. Suppose that
\begin{equation}\label{eq:open}
K_i\;\ge\;\E\big[(v_i-\delta)\one\{M_i<\delta\}\big]
\end{equation}
for every $\delta$ in a dense subset of $(r,\bar v]$. Then \eqref{eq:open} holds for every $\delta\in(r,\bar v]$, and
\begin{equation}\label{eq:closed}
K_i\;\ge\;\E\big[(v_i-\delta)\one\{M_i\le\delta\}\big]\qquad\text{for every }\delta\in[r,\bar v].
\end{equation}
The bids $\delta\in(r,\bar v]$ that are not atoms of the law of $M_i$ form a dense set, because a probability law has at most countably many atoms. In a BCCE the hypothesis holds on this set, since such a deviation ties with probability zero and wins exactly when $M_i<\delta$. If all buyers have the common value $v$ and submit a common bid $c$, so that $M_i=c$, then every deviation to $\delta\in(r,\bar v]$ earns at most $K_i$, and without a reserve so does the deviation to $0$.
\end{lemmaS}

\begin{proof}
For $\delta\in(r,\bar v]$ take $\delta_n\uparrow\delta$ in the dense set, and for $\delta\in[r,\bar v)$ take $\delta_n\downarrow\delta$. The indicators $\one\{M_i<\delta_n\}$ converge to $\one\{M_i<\delta\}$ in the first case and to $\one\{M_i\le\delta\}$ in the second, so dominated convergence gives \eqref{eq:open} and \eqref{eq:closed} on these ranges. At $\delta=\bar v$, \eqref{eq:closed} follows from \eqref{eq:open} because $v_i-\bar v\le0$. Under identical play, a deviator bidding $\delta\in(r,\bar v]$ wins outright when $c<\delta$ and shares with the $N-1$ other buyers when $c=\delta$, so the deviation payoff is a convex combination of the right sides of \eqref{eq:open} and \eqref{eq:closed}, hence at most $K_i$. Without a reserve, the deviation to $0$ earns $\frac1N\E[v\one\{c=0\}]$, which is at most the right side of \eqref{eq:closed} at $\delta=0$ because $v\ge0$.
\end{proof}

\section{Rate of Convergence}\label{app:rate}

\begin{lemmaS}[The interim guarantee converges at rate $1/N$]\label{lemma:interim_rate}
For every $N\ge2$, $0\le R_\infty-RG_N\le \dfrac{2\bar v}{N-1}$.
\end{lemmaS}

\begin{proof}
Write $\alpha=\tfrac{N-1}{N}$. Substituting $t=uF(v)$ in the two conditional means gives
\[
G(v)=\int_0^1F^{-1}\big(uF(v)\big)\,du,
\qquad
\beta_N(v)=\int_0^1F^{-1}\big(uF(v)\big)\,\alpha u^{\alpha-1}\,du,
\]
so $G(v)-\beta_N(v)=\int_0^1F^{-1}(uF(v))\big(1-\alpha u^{\alpha-1}\big)du$. The weight $1-\alpha u^{\alpha-1}$ is increasing in $u$, integrates to zero, is negative for $u<u_0\coloneqq(1-\tfrac1N)^N$ and positive above, and $F^{-1}(uF(v))$ is nondecreasing in $u$, so the integral is nonnegative by the Chebyshev integral inequality: $\beta_N\le G$ pointwise. For the upper bound, $|G-\beta_N|\le\bar v\int_0^1|1-\alpha u^{\alpha-1}|du=2\bar v\,(u_0^\alpha-u_0)=2\bar v\,\tfrac{u_0}{N-1}\le\tfrac{2\bar v}{N-1}$, using $u_0^{\alpha-1}=(1-\tfrac1N)^{-1}$. Integrating against $f$ gives the claim.
\end{proof}

\begin{proof}[Proof of Proposition \ref{prop:rate}]
The bound $0\le R_\infty-RG_N\le 2\bar v/(N-1)$ is Lemma \ref{lemma:interim_rate}. Write $K_N\coloneqq K^*_N$ for the rent at market size $N$ and $q_N\coloneqq q_{K_N}$ for the pooling quantile, so that $\Phi_\mu(q_N)=K_N$. The proof of Theorem \ref{thm:asymptotic} gives $R_\infty=\int_0^1\frac{\Phi_\mu(q)}{q}dq$ and $K_N\le m/N\to0$, hence $q_N\to0$.

\emph{Step 1: an exact decomposition.} Since $\Phi_\mu(q)\ge K_N$ exactly when $q\ge q_N$, \eqref{eq:Rcal} reads $CRG_N=\int_{q_N}^1\frac{\Phi_\mu(q)-K_N}{q}dq$, and subtracting from $R_\infty$,
\[
R_\infty-CRG_N\;=\;T_N+K_N\log\frac{1}{q_N},\qquad T_N\coloneqq\int_0^{q_N}\frac{\Phi_\mu(q)}{q}\,dq .
\]
The first term is the limiting revenue on the pooled quantiles, and the second is the rent on the bidding quantiles.

\emph{Step 2: the order of the rent.} By \eqref{eq:fixed_point}, $NK_N=m-CRG_N\to m-R_\infty$, and this limit is strictly positive because $m-R_\infty=g_N(0)>0$ by Lemma \ref{lemma:concavity}. Hence $K_N=\frac{m-R_\infty}{N}(1+o(1))$ and $\log\frac1{K_N}=\log N+O(1)$.

\emph{Step 3: the order of the pooling quantile.} This is the only step that uses the continuity and positivity of $f$ at $0$ in Assumption \ref{ass:standing}. They give $F(x)=f(0)x(1+o(1))$ as $x\downarrow0$, hence $F^{-1}(t)=\frac{t}{f(0)}(1+o(1))$ as $t\downarrow0$ and, integrating, $\Phi_\mu(q)=\frac{q^2}{2f(0)}(1+o(1))$ as $q\downarrow0$. At $q=q_N$ this reads $K_N=\frac{q_N^2}{2f(0)}(1+o(1))$, so
\[
\log\frac1{q_N}\;=\;\frac12\log\frac1{K_N}+O(1)\;=\;\frac12\log N+O(1).
\]
The square root in $q_N=\sqrt{2f(0)K_N}\,(1+o(1))$ is the source of the factor $\tfrac12$ in the proposition.

\emph{Step 4: the pooled term is of smaller order.} Since $\Phi_\mu(q)/q\le F^{-1}(q)$, $0\le T_N\le\int_0^{q_N}F^{-1}(q)\,dq=\Phi_\mu(q_N)=K_N=O(1/N)$.

\emph{Step 5: conclusion.} By Steps 2 and 3, $K_N\log\frac1{q_N}=\frac{m-R_\infty}{N}(1+o(1))\big(\tfrac12\log N+O(1)\big)=\frac{(m-R_\infty)\log N}{2N}(1+o(1))$, and adding $T_N=O(1/N)=o(\log N/N)$ gives the expansion for $R_\infty-CRG_N$. Finally $RG_N-CRG_N=(R_\infty-CRG_N)-(R_\infty-RG_N)$, and the second bracket is $O(1/N)=o(\log N/N)$ by Lemma \ref{lemma:interim_rate}, so $RG_N-CRG_N$ has the same expansion.
\end{proof}

\section{Convex Order and the Binary Bound}\label{app:binary}

\begin{proof}[Proof of Proposition \ref{prop:convex_order}]
Suppose $\mu\cx\nu$. By the quantile characterization of the convex order \citep[Theorem 3.A.5]{ShakedShanthikumar2007}, the two distributions share the mean $m$ and satisfy $\Phi_\nu(q)\le\Phi_\mu(q)$ for all $q$. Hence $\mathcal{R}_\nu\le\mathcal{R}_\mu$ pointwise by \eqref{eq:Rcal}, so
\[
g^\nu_N(K)=m-NK-\mathcal{R}_\nu(K)\;\ge\;m-NK-\mathcal{R}_\mu(K)=g^\mu_N(K)\qquad\text{for all }K .
\]
If $\nu=\mu$ there is nothing to prove, and otherwise $\nu$ is non-degenerate. By Lemma \ref{lemma:concavity}, $g^\nu_N$ is nonnegative exactly on $[0,K^*_\nu]$, where $K^*_\nu$ is its unique zero, and $CRG(\nu)=m-NK^*_\nu$ by Theorem \ref{thm:general}. The function $g^\mu_N$ is also concave, with $g^\mu_N(0)\ge0$, so it is nonnegative exactly on $[0,K^*_\mu]$, where $K^*_\mu$ is its largest zero, and Step 1 of Appendix \ref{app:theorem1} gives $CRG(\mu)\ge m-NK^*_\mu$. This also covers a degenerate $\mu$, to which the uniqueness part of Lemma \ref{lemma:concavity} does not apply. Since $g^\nu_N(K^*_\mu)\ge g^\mu_N(K^*_\mu)=0$, we have $K^*_\mu\le K^*_\nu$, and therefore $CRG(\nu)=m-NK^*_\nu\le m-NK^*_\mu\le CRG(\mu)$.

For the second claim, among distributions supported on $[0,\bar v]$ with mean $E$, the binary distribution $\underline\mu$ on $\{0,\bar v\}$ is the largest in the convex order: for any such $\mu$ and any convex $\phi$, $\int\phi\,d\mu\le\int\phi\,d\underline\mu$, because on $[0,\bar v]$ a convex function lies below the line connecting $(0,\phi(0))$ and $(\bar v,\phi(\bar v))$. Hence $CRG(\underline\mu)\le CRG(\mu)$ for every such $\mu$.
\end{proof}

\begin{proof}[Proof of Corollary \ref{cor:binary}]
Write $\Phi$ and $\mathcal R$ for $\Phi_{\underline\mu}$ and $\mathcal R_{\underline\mu}$. For $\underline\mu$ we have $F^{-1}(t)=0$ for $t<p_0$ and $F^{-1}(t)=\bar v$ for $t>p_0$, so $\Phi(q)=\bar v(q-p_0)^+$ and $q_K=p_0+K/\bar v$. Then
\[
\mathcal{R}(K)=\int_{q_K}^{1}\frac{\bar v(q-p_0)-K}{q}\,dq=\bar v\int_{q_K}^1\frac{q-q_K}{q}\,dq=\bar v\big[(1-q_K)+q_K\log q_K\big].
\]
By Theorem \ref{thm:general}, $CRG(\underline\mu)=E-NK^*$, where $K^*$ solves \eqref{eq:fixed_point} with $m=E$. Substituting $K=\bar v(q_K-p_0)$ and $E=\bar v(1-p_0)$ into \eqref{eq:fixed_point} and simplifying yields $(N-1)(p_0-q_K)=q_K\log q_K$, so $\hat q=q_{K^*}$ and $CRG=E-N\bar v(\hat q-p_0)$. Uniqueness of the root is Lemma \ref{lemma:concavity}. As $N\to\infty$ we have $K^*\to0$ and $\hat q\to p_0$, so
\[
CRG\;\longrightarrow\;\bar v\big[(1-p_0)+p_0\log p_0\big]=E+(\bar v-E)\log\Big(1-\frac{E}{\bar v}\Big). \qedhere
\]
\end{proof}

\end{document}